\documentclass[showpacs,preprintnumbers,twocolumn,superscriptaddress,aps,nofootinbib]{revtex4-1}
\usepackage{amssymb}
\usepackage[centertags]{amsmath}
\usepackage{txfonts}
\usepackage{epsfig}
\usepackage{bm}
\usepackage{color}
\usepackage{graphicx,graphics}
\usepackage{multirow}
\usepackage{float}
\usepackage{ulem}
\usepackage{appendix}
\usepackage[pdfstartview=FitH]{hyperref}
\hypersetup{colorlinks=true, citecolor=blue,linkcolor=blue,filecolor=black,urlcolor=blue}  
\allowdisplaybreaks[2]

\begin{document}

\title{Production characteristics of light nuclei, hypertritons and $\Omega$-hypernuclei in Pb+Pb collisions at $\sqrt{s_{NN}}=5.02$ TeV}

\author{Rui-Qin Wang}
\email {wangrq@qfnu.edu.cn}
\affiliation{School of Physics and Physical Engineering, Qufu Normal University, Shandong 273165, China}

\author{Xin-Lei Hou}
\affiliation{School of Physics and Physical Engineering, Qufu Normal University, Shandong 273165, China}

\author{Yan-Hao Li}
\affiliation{School of Physics and Physical Engineering, Qufu Normal University, Shandong 273165, China}

\author{Jun Song}
\email {songjun2011@jnxy.edu.cn}
\affiliation{School of Physical Science and Intelligent Engineering, Jining University, Shandong 273155, China}

\author{Feng-Lan Shao}
\email {shaofl@mail.sdu.edu.cn}
\affiliation{School of Physics and Physical Engineering, Qufu Normal University, Shandong 273165, China}

\begin{abstract}

We extend an analytical nucleon coalescence model with hyperons to study productions of light nuclei, hypertritons and $\Omega$-hypernuclei in Pb+Pb collisions at $\sqrt{s_{NN}}=5.02$ TeV.
We derive the formula of the momentum distribution of two bodies coalescing into dibaryon states and that of three bodies coalescing into tribaryon states.
We explain the available data of the coalescence factors $B_2$ and $B_3$, the transverse momentum spectra, the averaged transverse momenta, the yield rapidity densities, yield ratios of the deuteron, antihelium-3, antitriton, hypertriton measured by the ALICE collaboration, and give predictions of different $\Omega$-hypernuclei, e.g., $H(p\Omega^-)$, $H(n\Omega^-)$ and $H(pn\Omega^-)$.
We find two groups of interesting observables, the averaged transverse momentum ratios of light (hyper-)nuclei to protons (hyperons) and the centrality-dependent yield ratios of theirs.
The former group exhibits a reverse-hierarchy of the nucleus size, and the latter is helpful for the judgements of the nucleus production mechanism as well as the nucleus own size.

\end{abstract}


\maketitle

\section{Introduction}   \label{introduction}

In ultra-relativistic heavy ion collisions, light nuclei and hypernuclei such as the deuteron ($d$), helium-3 ($^3$He), triton ($t$), hypertriton ($^3_{\Lambda}$H) and helium-4 ($^4$He) are a special group of observerables~\cite{Chen:2018tnh,Gutbrod:1976zzr,Aichelin:1991xy,Andronic:2017pug,Bzdak:2019pkr,Sun:2017xrx,Sun:2018jhg,Luo:2020pef,Steinheimer:2008hr,NPLQCD:2012mex,Nemura:1999qp,Junnarkar:2019equ,Morita:2019rph}.
They are composite clusters and their production mechanism is currently still under debate.
The productions of such composite objects closely relate with many fundamental issues in relativistic heavy ion collision physics, e.g., the hadronization mechanism~\cite{Aichelin:1991xy}, the structure of the quantum chromodynamics phase diagram~\cite{Andronic:2017pug,Sun:2017xrx,Sun:2018jhg,Bzdak:2019pkr,Luo:2020pef}, the local baryon-strangeness correlation~\cite{Steinheimer:2008hr}, the hyperon-nucleon interaction~\cite{NPLQCD:2012mex,Nemura:1999qp} and the search of more hadronic molecular states~\cite{Junnarkar:2019equ,Morita:2019rph}.

In recent decades, the production of light nuclei and hyper-nuclei in ultra-relativistic heavy ion collisions has always attracted much attention both in experiment~\cite{PHENIX:2007tef,NA49:2016qvu,STAR:2001pbk,STAR:2016ydv,STAR:2020hya,STAR:2019sjh,STAR:2022hbp,ALICE:2020chv,ALICE:2022veq,STAR:2019wjm,STAR:2021orx,ALICE:2022sco,ALICE:2015oer} and in theory~\cite{Braun-Munzinger:2018hat,Oliinychenko:2020ply,Mrowczynski:2020ugu,Donigus:2022haq,Dover:1991zn,Chen:2003qj,Andronic:2010qu}.
The STAR experiment at the BNL Relativistic Heavy Ion Collider (RHIC) and the ALICE experiment at the CERN Large Hadron Collider (LHC) have put many efforts on measurements of 
light nuclei~\cite{ALICE:2020chv,STAR:2016ydv,STAR:2020hya,STAR:2019sjh,STAR:2022hbp,ALICE:2022veq} and hypernuclei~\cite{STAR:2019wjm,STAR:2021orx,ALICE:2022sco,ALICE:2015oer}.
In theory two production mechanisms, the thermal production mechanism~\cite{Mekjian:1977ei,Siemens:1979dz,Andronic:2010qu,Cleymans:2011pe,Cai:2019jtk} and the coalescence mechanism~\cite{Schwarzschild:1963zz,Sato:1981ez,Dover:1991zn,Mattiello:1995xg,Nagle:1996vp,Mattiello:1996gq,Chen:2003qj,Polleri:1997bp,Scheibl:1998tk,Sharma:2018dyb,Zhao:2018lyf}, have proved to be successful in describing formations of such composite objects. 

The coalescence mechanism, in which light nuclei and hypernuclei are assumed to be produced by the coalescence of the jacent nucleons and hyperons in the phase space, possesses some unique characteristics.
In order to see whether, if so, to what extent, these characteristics depend on the particular coalescence models used in obtaining these characteristics, 
we in our previous works~\cite{Zhao:2022xkz,Wang:2022hja,Wang:2020zaw,Wang:2023rpd} developed an analytical description for the production of different species of light nuclei.
We applied the analytical nucleon coalescence model to Au+Au collisions at the RHIC to successfully explain the energy-dependent behaviors of $d$, $t$, $^3$He and $^4$He~\cite{Zhao:2022xkz,Wang:2022hja}.
We also applied it to pp, p+Pb and Pb+Pb collisions at the LHC to understand the different behaviors of the coalescence factors $B_2$ and $B_3$~\cite{Wang:2020zaw} from small to large collision systems, and gave a series of concise production correlations of $d$, $^3$He and $t$~\cite{Wang:2023rpd}.

Very recently, the ALICE collaboration published the most precise measurements to date of $d$, $^3$He, $t$ and especially $^3_{\Lambda}$H in Pb+Pb collisions at $\sqrt{s_{NN}}=5.02$ TeV~\cite{ALICE:2022veq,ALICE:2022sco,ALICE:2024koa}. 
In this work, we extend the coalescence model considering the coordinate-momentum correlation~\cite{Wang:2023rpd} to include the hyperon coalescence besides the nucleon coalescence, and apply it to simultaneously study productions of light nuclei, the $^3_{\Lambda}$H and different $\Omega$-hypernuclei. 
One main goal of this article is to give an overall comprehension for the newest data in Pb+Pb collisions with the highest collision energy up to now.
The other goal is to bring production characteristics of light nuclei and hypernuclei originating from the coalescence itself to light.

The paper is organized as follows. 
In Sec.~\ref{model}, we give an introduction to the coalescence model. 
We present the formula of the momentum distribution of two baryons coalescing into dibaryon states and that of three baryons coalescing into tribaryon states.
In Sec.~\ref{lightnuclei}, we study behaviors of the coalescence factors $B_2$ and $B_3$ as functions of the collision centrality and the transverse momentum per nucleon.
We also study the transverse momentum ($p_T$) spectra, the averaged transverse momenta $\langle p_T\rangle$, the yield rapidity densities $dN/dy$ and yield ratios of $d$, $^3\overline{\text{He}}$ and $\bar t$.
In Sec.~\ref{hynucleiResults}, we present results of the $^3_{\Lambda}$H and $\Omega$-hypernuclei.
We specially study the averaged transverse momentum ratios ${\langle p_T\rangle_d}/{\langle p_T\rangle_p}$, ${\langle p_T\rangle_{H(p\Omega^-)}}/{\langle p_T\rangle_{\Omega^-}}$, ${\langle p_T\rangle_{H(n\Omega^-)}}/{\langle p_T\rangle_{\Omega^-}}$, 
${\langle p_T\rangle_t}/{\langle p_T\rangle_p}$, ${\langle p_T\rangle_{\mathrm{^3He}}}/{\langle p_T\rangle_p}$, ${\langle p_T\rangle_{^3_{\Lambda}\mathrm{H}}}/{\langle p_T\rangle_{\Lambda}}$, ${\langle p_T\rangle_{H(pn\Omega^-)}}/{\langle p_T\rangle_{\Omega^-}}$,
and centrality-dependent behaviors of yield ratios $d/p$, $H(p\Omega^-)/\Omega^-$, $H(n\Omega^-)/\Omega^-$, $t/p$, $^3\mathrm{He}/p$, $^3_{\Lambda}\mathrm{H}/\Lambda$, $H(pn\Omega^-)/\Omega^-$. 
In Sec.~\ref{summary}, we give our summary.

\section{The coalescence model}   \label{model}

In this section we extend the analytical nucleon coalescence model in our previous work~\cite{Wang:2023rpd} to include the coalescence of hyperons.
In the current coalescence model, the coalescence process is executed on an equivalent kinetic freeze-out surface formed from different times.
To make the analytical and intuitive insights possible, we abandon carrying out the time evolution step by step but absorb the finite emission duration in an effective volume parameter.
We first present the formalism of two baryons coalescing into $d$-like dibaryon states. 
We then give analytical expressions of three baryons coalescing into $^3$He, $t$ and their partners in the strange sector.

\subsection{Formalism of two bodies coalescing into dibaryon states}      \label{2hcoHj}

We begin with a hadronic system produced at the final stage of the evolution of high energy collision and suppose the dibaryon state $H_{j}$ is formed via the coalescence of two baryons $h_1$ and $h_2$.
We use $f_{H_{j}}(\bm{p})$ to denote the three-dimensional momentum distribution of the produced $H_{j}$ and it is given by
{\setlength\arraycolsep{0pt}
\begin{eqnarray}
f_{H_{j}}(\bm{p}) =&&  \int d\bm{x}_1d\bm{x}_2 d\bm{p}_1 d\bm{p}_2  f_{h_1h_2}(\bm{x}_1,\bm{x}_2;\bm{p}_1,\bm{p}_2)   \nonumber  \\
&& ~~~~\times  \mathcal {R}_{H_{j}}(\bm{x}_1,\bm{x}_2;\bm{p}_1,\bm{p}_2,\bm{p}),      \label{eq:fHj2hgeneral} 
\end{eqnarray} }%
where $f_{h_1h_2}(\bm{x}_1,\bm{x}_2;\bm{p}_1,\bm{p}_2)$ is two-baryon joint coordinate-momentum distribution; 
$\mathcal {R}_{H_{j}}(\bm{x}_1,\bm{x}_2;\bm{p}_1,\bm{p}_2,\bm{p})$ is the kernel function of the $H_{j}$.
Here and from now on we use bold symbols to denote three-dimensional coordinate or momentum vectors.

In terms of the normalized joint coordinate-momentum distribution denoted by the superscript `$(n)$', we have
{\setlength\arraycolsep{0pt}
\begin{eqnarray}
f_{H_{j}}(\bm{p})=&& N_{h_1h_2} \int d\bm{x}_1d\bm{x}_2 d\bm{p}_1 d\bm{p}_2  f^{(n)}_{h_1h_2}(\bm{x}_1,\bm{x}_2;\bm{p}_1,\bm{p}_2)    \nonumber  \\
&& ~~~~~~~~~~~~\times   \mathcal {R}_{H_{j}}(\bm{x}_1,\bm{x}_2;\bm{p}_1,\bm{p}_2,\bm{p}).      \label{eq:fHj2hgeneral1}               
\end{eqnarray} }%
$N_{h_1h_2}=N_{h_1}N_{h_2}$ is the number of all possible $h_1h_2$-pairs in the considered hadronic system, and $N_{h_i}~ (i=1,2)$ is the number of the baryons $h_i$.

The kernel function $\mathcal {R}_{H_{j}}(\bm{x}_1,\bm{x}_2;\bm{p}_1,\bm{p}_2,\bm{p})$ denotes the probability density for $h_1$, $h_2$ with
momenta $\bm{p}_1$ and $\bm{p}_2$ at $\bm{x}_1$ and $\bm{x}_2$ to combine into an $H_{j}$ of momentum $\bm{p}$.
It carries the kinetic and dynamical information of $h_1$ and $h_2$ combining into $H_{j}$,
and its precise expression should be constrained by such as the momentum conservation, constraints due to intrinsic quantum numbers e.g., spin~\cite{Wang:2020zaw,Zhao:2022xkz,Wang:2022hja,Wang:2023rpd}.
To take these constraints into account explicitly, we rewrite the kernel function in the following form
{\setlength\arraycolsep{0pt}
\begin{eqnarray}
\mathcal {R}_{H_{j}}(\bm{x}_1,\bm{x}_2;\bm{p}_1,\bm{p}_2,\bm{p}) = g_{H_{j}} \mathcal {R}_{H_{j}}^{(x,p)}(\bm{x}_1,\bm{x}_2;\bm{p}_1,\bm{p}_2)  \delta(\displaystyle{\sum^2_{i=1}} \bm{p}_i-\bm{p}).~~~~   \label{eq:RHj2h}  
\end{eqnarray} }%
The spin degeneracy factor $g_{H_{j}} = (2J_{H_{j}}+1) /[\prod \limits_{i=1}^2(2J_{h_i}+1)]$, where $J_{H_{j}}$ is the spin of the produced $H_{j}$ and $J_{h_i}$ is that of the primordial baryon $h_i$. 
The Dirac $\delta$ function guarantees the momentum conservation in the coalescence process.
The remaining $\mathcal {R}_{H_{j}}^{(x,p)}(\bm{x}_1,\bm{x}_2;\bm{p}_1,\bm{p}_2)$ can be solved from the Wigner transformation as the $H_{j}$ wave function is given.
Considering the wave function of a spherical harmonic oscillator is particularly tractable and useful for analytical insight, we adopt this profile as in Refs.~\cite{Chen:2003ava,Zhu:2015voa} and have
{\setlength\arraycolsep{0pt}
\begin{eqnarray}
 \mathcal {R}^{(x,p)}_{H_{j}}(\bm{x}_1,\bm{x}_2;\bm{p}_1,\bm{p}_2) = 8e^{-\frac{(\bm{x}'_1-\bm{x}'_2)^2}{2\sigma^2}}
	e^{-\frac{2\sigma^2(m_2\bm{p}'_{1}-m_1\bm{p}'_{2})^2}{(m_1+m_2)^2\hbar^2c^2}}.      \label{eq:RHj2hxp} 
\end{eqnarray} }%
The superscript `$'$' in the coordinate or momentum variable denotes the baryon coordinate or momentum in the rest frame of the $h_1h_2$-pair.
$m_1$ and $m_2$ are the mass of $h_1$ and that of $h_2$.
The width parameter $\sigma=\sqrt{\frac{2(m_1+m_2)^2}{3(m_1^2+m_2^2)}} R_{H_{j}}$, where $R_{H_{j}}$ is the root-mean-square radius of $H_{j}$.
The factor $\hbar c$ comes from the used GeVfm unit, and it is 0.197 GeVfm.

Substituting Eqs.~(\ref{eq:RHj2h}) and (\ref{eq:RHj2hxp}) into Eq.~(\ref{eq:fHj2hgeneral1}), we have
{\setlength\arraycolsep{0.2pt}
\begin{eqnarray}
 f_{H_{j}}(\bm{p})=&& N_{h_1h_2} g_{H_{j}} \int d\bm{x}_1d\bm{x}_2d\bm{p}_1d\bm{p}_2 f^{(n)}_{h_1h_2}(\bm{x}_1,\bm{x}_2;\bm{p}_1,\bm{p}_2)    \nonumber  \\
&& \times
8e^{-\frac{(\bm{x}'_1-\bm{x}'_2)^2}{2\sigma^2}}   e^{-\frac{2\sigma^2(m_2\bm{p}'_{1}-m_1\bm{p}'_{2})^2}{(m_1+m_2)^2\hbar^2c^2}}
	\delta(\displaystyle{\sum^2_{i=1}} \bm{p}_i-\bm{p}).    \label{eq:fHj2h}  
\end{eqnarray} }%
This is the general formalism of the $H_j$ produced via the coalescence of two baryons $h_1$ and $h_2$.

Noticing that the root-mean-square radius $R_{H_{j}}$ of the $d$-like dibaryon state $H_{j}$ is always considered to be about or larger than 2 fm, $\sigma$ is even larger than $R_{H_{j}}$.
So the gaussian width in the momentum-dependent part of the kernel function in Eq.~(\ref{eq:fHj2h}) has a small value, about or smaller than 0.1 GeV.
Therefore, we approximate the gaussian form of the momentum-dependent kernel function to be a $\delta$ function form as follows
{\setlength\arraycolsep{0.2pt}
\begin{equation}
e^{-\frac{(\bm{p}'_{1}-\frac{m_1}{m_2}\bm{p}'_{2})^2} {(1+\frac{m_1}{m_2})^2 \frac{\hbar^2c^2}{2\sigma^2}}} \approx
\left[ \frac{\hbar c}{\sigma} (1+\frac{m_1}{m_2}) \sqrt{\frac{\pi}{2}} \right]^3 \delta(\bm{p}'_{1}-\frac{m_1}{m_2}\bm{p}'_{2}).   \label{eq:deltaapp}  
\end{equation}
Substituting Eq.~(\ref{eq:deltaapp}) into Eq.~(\ref{eq:fHj2h}) and integrating $\bm{p}_1$ and $\bm{p}_2$, we can obtain
{\setlength\arraycolsep{0.2pt}
\begin{eqnarray}
&& f_{H_{j}}(\bm{p})=8 g_{H_{j}} N_{h_1h_2} \int d\bm{x}_1d\bm{x}_2d\bm{p}_1d\bm{p}_2 f^{(n)}_{h_1h_2}(\bm{x}_1,\bm{x}_2;\bm{p}_1,\bm{p}_2)    \nonumber  \\
&&~~ \times
e^{-\frac{(\bm{x}'_1-\bm{x}'_2)^2}{2 \sigma^2}}   (\frac{\hbar c\sqrt{\pi}}{\sqrt{2} \sigma})^3 (1+\frac{m_1}{m_2})^3
\delta(\bm{p}'_{1}-\frac{m_1}{m_2} \bm{p}'_{2})  \delta(\displaystyle{\sum^2_{i=1}} \bm{p}_i-\bm{p})                 \nonumber    \\                             
&&= 8 g_{H_{j}} N_{h_1h_2} \int d\bm{x}_1d\bm{x}_2d\bm{p}_1d\bm{p}_2 f^{(n)}_{h_1h_2}(\bm{x}_1,\bm{x}_2;\bm{p}_1,\bm{p}_2)     \nonumber  \\
&&~~ \times  e^{-\frac{(\bm{x}'_1-\bm{x}'_2)^2}{2 \sigma^2}}  
(\frac{\hbar c\sqrt{\pi}}{\sqrt{2} \sigma})^3 (1+\frac{m_1}{m_2})^3
\gamma \delta(\bm{p}_{1}-\frac{m_1}{m_2} \bm{p}_{2})  \delta(\displaystyle{\sum^2_{i=1}} \bm{p}_i-\bm{p})         \nonumber   \\              
&&=8g_{H_{j}} N_{h_1h_2} \gamma (\frac{\hbar c\sqrt{\pi}}{\sqrt{2} \sigma})^3     \nonumber  \\
&&~~ \times  \int d\bm{x}_1d\bm{x}_2      f^{(n)}_{h_1h_2}(\bm{x}_1,\bm{x}_2;\frac{m_1 \bm{p}}{m_1+m_2},\frac{m_2 \bm{p}}{m_1+m_2})  e^{-\frac{(\bm{x}'_1-\bm{x}'_2)^2}{2 \sigma^2}},    \label{eq:fH-p}   
\end{eqnarray}	
where $\gamma$ comes from the Lorentz transformation $\bm{p}'_{1}-\frac{m_1}{m_2}\bm{p}'_{2}=\frac{1}{\gamma} (\bm{p}_{1}-\frac{m_1}{m_2}\bm{p}_{2})$.   

Changing coordinate variables in Eq.~(\ref{eq:fH-p}) to be $\bm{X}=\frac{\bm{x}_1+\bm{x}_2}{\sqrt{2}} $ and $\bm{r}=\frac{\bm{x}_1-\bm{x}_2} {\sqrt{2}}$, we have  
{\setlength\arraycolsep{0pt}
\begin{eqnarray}
&& f_{H_{j}}(\bm{p})= 8g_{H_{j}} N_{h_1h_2}  (\frac{\hbar c\sqrt{\pi}}{\sqrt{2} \sigma})^3 \gamma     \nonumber  \\
&&~~~~~~ \times  \int d\bm{X}d\bm{r}  f^{(n)}_{h_1h_2}(\bm{X},\bm{r};\frac{m_1}{m_1+m_2} \bm{p},\frac{m_2}{m_1+m_2} \bm{p})    e^{-\frac{\bm{r}'^2}{\sigma^2}}.   \label{eq:fH-Xr}  
\end{eqnarray} }%
Considering the strong interaction and the coalescence are local, we neglect the effect of collective motion on the center of mass coordinate and assume it is factorized, i.e.,
{\setlength\arraycolsep{0pt}
\begin{eqnarray}
&&   f^{(n)}_{h_1h_2}(\bm{X},\bm{r};\frac{m_1}{m_1+m_2} \bm{p},\frac{m_2}{m_1+m_2} \bm{p}) =  f^{(n)}_{h_1h_2}(\bm{X})    \nonumber  \\
&&~~~~~~~~~~~~~~~~~~~~~~~~~~~~~ \times f^{(n)}_{h_1h_2}(\bm{r};\frac{m_1}{m_1+m_2} \bm{p},\frac{m_2}{m_1+m_2} \bm{p}).  \label{eq:fh1h2-Xr}  
\end{eqnarray} }%
Substituting Eq.~(\ref{eq:fh1h2-Xr}) into Eq.~(\ref{eq:fH-Xr}), we have
{\setlength\arraycolsep{0pt}
\begin{eqnarray}
&& f_{H_{j}}(\bm{p})= 8g_{H_{j}} N_{h_1h_2}  (\frac{\hbar c\sqrt{\pi}}{\sqrt{2} \sigma})^3 \gamma     \nonumber  \\
&&~~~~~~~~~~~~~~ \times  \int d\bm{r} f^{(n)}_{h_1h_2}(\bm{r};\frac{m_1}{m_1+m_2} \bm{p},\frac{m_2}{m_1+m_2} \bm{p})    e^{-\frac{\bm{r}'^2}{\sigma^2}}.   \label{eq:fH-r} 
\end{eqnarray} }%

We adopt the frequently-used gaussian form for the relative coordinate distribution as in such as Ref.~\cite{Mrowczynski:2016xqm}, i.e.,
{\setlength\arraycolsep{0pt}
\begin{eqnarray}
&& f^{(n)}_{h_1h_2}(\bm{r};\frac{m_1}{m_1+m_2} \bm{p},\frac{m_2}{m_1+m_2} \bm{p}) =  \frac{1}{\left[\pi C_w R_f^2(\bm{p})\right]^{3/2}} 
  e^{-\frac{\bm{r}^2}{C_w R_f^2(\bm{p})}}     \nonumber  \\
&&~~~~~~~~~~~~~~~~~~~~~~~ \times f^{(n)}_{h_1h_2}(\frac{m_1}{m_1+m_2} \bm{p},\frac{m_2}{m_1+m_2} \bm{p}).     \label{eq:fh1h2-rp}    
\end{eqnarray} }%
Here $R_f(\bm{p})$ is the effective radius of the hadronic source system at the $H_j$ freeze-out.
$C_w$ is a distribution width parameter and it is 2~\cite{Mrowczynski:2016xqm}. 
Gaussian profile of the relative coordinate distribution is convenient for analytical calculations, and there are some empirical arguments in favor of such choice~\cite{Kisiel:2006is,NA49:2008ofa}.

With instantaneous coalescence in the rest frame of $h_1h_2$-pair, i.e., $\Delta t'=0$, we get the Lorentz transformation
\begin{eqnarray}
\bm{r} = \bm{r}' +(\gamma-1)\frac{\bm{r}'\cdot \bm{\beta}}{\beta^2}\bm{\beta},    \label{eq:LorentzTr}  
\end{eqnarray}%
where $\bm{\beta}$ is the three-dimensional velocity vector of the center-of-mass frame of $h_1h_2$-pair in the laboratory frame and $\gamma$ is the corresponding contraction factor.
Substituting Eq.~(\ref{eq:fh1h2-rp}) into Eq.~(\ref{eq:fH-r}) and using Eq.~(\ref{eq:LorentzTr}) to integrate from the relative coordinate variable, we obtain 
{\setlength\arraycolsep{0pt}
\begin{eqnarray}
&&f_{H_{j}}(\bm{p}) = \frac{ (\sqrt{2\pi}\hbar c)^3 g_{H_{j}} \gamma}{\left[C_w R_f^2(\bm{p})+\sigma^2\right] \sqrt{C_w [R_f(\bm{p})/\gamma]^2+\sigma^2}}    \nonumber  \\
&&~~~~~~~~~~~~~~ \times  f_{h_1h_2}(\frac{m_1}{m_1+m_2} \bm{p},\frac{m_2}{m_1+m_2} \bm{p}).   \label{eq:fHjh1h2} 
\end{eqnarray} }%
Ignoring correlations between $h_1$ and $h_2$, we have the three-dimensional momentum distribution of the $H_j$ as
{\setlength\arraycolsep{0pt}
\begin{eqnarray}
f_{H_{j}}(\bm{p}) =&& \frac{ (\sqrt{2\pi}\hbar c)^3 g_{H_{j}} \gamma}{\left[C_w R_f^2(\bm{p})+\sigma^2\right] \sqrt{C_w [R_f(\bm{p})/\gamma]^2+\sigma^2}}    \nonumber  \\
&& \times  f_{h_1}(\frac{m_1\bm{p}}{m_1+m_2}) f_{h_2}(\frac{m_2\bm{p}}{m_1+m_2}).    \label{eq:fHj2h-approx}
\end{eqnarray} }%

Denoting the Lorentz invariant momentum distribution $\dfrac{d^{2}N}{2\pi p_{T}dp_{T}dy}$ with $f^{(inv)}$, we finally have
{\setlength\arraycolsep{0.2pt}
\begin{eqnarray}
&&f_{H_j}^{(inv)}(p_T,y) =\frac{ (\sqrt{2\pi}\hbar c)^3 g_{H_{j}} }{\left[C_w R_f^2(p_T,y)+\sigma^2\right] \sqrt{C_w [R_f(p_T,y)/\gamma]^2+\sigma^2}}      \nonumber  \\
&& ~~~~~ \times \frac{m_{H_j}}{m_1m_2} f_{h_1}^{(inv)}(\frac{m_1p_{T}}{m_1+m_2},y)  f_{h_2}^{(inv)}(\frac{m_2p_{T}}{m_1+m_2},y),     \label{eq:pt-Hj2h}
\end{eqnarray} }%
where $y$ is the longitudinal rapidity and $m_{H_j}$ is the mass of the $H_j$.

\subsection{Formalism of three bodies coalescing into tribaryon states}

For tribaryon state $H_{j}$ formed via the coalescence of three baryons $h_1$, $h_2$ and $h_3$, the momentum distribution $f_{H_{j}}(\bm{p})$ is
{\setlength\arraycolsep{0pt}
\begin{eqnarray}
&&  f_{H_{j}}(\bm{p})= N_{h_1h_2h_3}    \nonumber  \\
&&~~ \times \int d\bm{x}_1d\bm{x}_2d\bm{x}_3 d\bm{p}_1 d\bm{p}_2d\bm{p}_3  f^{(n)}_{h_1h_2h_3}(\bm{x}_1,\bm{x}_2,\bm{x}_3;\bm{p}_1,\bm{p}_2,\bm{p}_3)    \nonumber  \\
&&~~~~~~~~~~ \times \mathcal {R}_{H_{j}}(\bm{x}_1,\bm{x}_2,\bm{x}_3;\bm{p}_1,\bm{p}_2,\bm{p}_3,\bm{p}).     \label{eq:fHj3hgeneral1}
\end{eqnarray} }%
$N_{h_1h_2h_3}$ is the number of all possible $h_1h_2h_3$-clusters and it equals to $N_{h_1}N_{h_2}N_{h_3},~N_{h_1}(N_{h_1}-1)N_{h_3}$ for $h_1 \neq h_2 \neq h_3$, $h_1 = h_2 \neq h_3$, respectively.
$f^{(n)}_{h_1h_2h_3}(\bm{x}_1,\bm{x}_2,\bm{x}_3;\bm{p}_1,\bm{p}_2,\bm{p}_3)$ is the normalized three-baryon joint coordinate-momentum distribution.
$\mathcal {R}_{H_{j}}(\bm{x}_1,\bm{x}_2,\bm{x}_3;\bm{p}_1,\bm{p}_2,\bm{p}_3,\bm{p})$ is the kernel function.

We rewrite the kernel function as
{\setlength\arraycolsep{0pt}
\begin{eqnarray}
\mathcal {R}_{H_{j}}(\bm{x}_1,\bm{x}_2,\bm{x}_3;\bm{p}_1,\bm{p}_2,\bm{p}_3,\bm{p}) =&& g_{H_{j}}
	\mathcal {R}_{H_{j}}^{(x,p)}(\bm{x}_1,\bm{x}_2,\bm{x}_3;\bm{p}_1,\bm{p}_2,\bm{p}_3)     \nonumber  \\
&& \times  \delta(\displaystyle{\sum^3_{i=1}} \bm{p}_i-\bm{p}).   \label{eq:RHj3h}  
\end{eqnarray} }%
The spin degeneracy factor $g_{H_{j}} = (2J_{H_{j}}+1) /[\prod \limits_{i=1}^3(2J_{h_i}+1)]$.
The Dirac $\delta$ function guarantees the momentum conservation.
$\mathcal {R}_{H_{j}}^{(x,p)}(\bm{x}_1,\bm{x}_2,\bm{x}_3;\bm{p}_1,\bm{p}_2,\bm{p}_3)$ solving from the Wigner transformation~\cite{Chen:2003ava,Zhu:2015voa} is
{\setlength\arraycolsep{0pt}
\begin{eqnarray}
&& \mathcal {R}^{(x,p)}_{H_{j}}(\bm{x}_1,\bm{x}_2,\bm{x}_3;\bm{p}_1,\bm{p}_2,\bm{p}_3) = 8^2 e^{-\frac{(\bm{x}'_1-\bm{x}'_2)^2}{2\sigma_1^2}}
	e^{-\frac{2(\frac{m_1\bm{x}'_1}{m_1+m_2}+\frac{m_2\bm{x}'_2}{m_1+m_2}-\bm{x}'_3)^2}{3\sigma_2^2}}    \nonumber  \\
&&~~~~~~~~~~~~~~~~~~~~~~~~~~ \times e^{-\frac{2\sigma_1^2(m_2\bm{p}'_{1}-m_1\bm{p}'_{2})^2}{(m_1+m_2)^2\hbar^2c^2}}
	e^{-\frac{3\sigma_2^2[m_3\bm{p}'_{1}+m_3\bm{p}'_{2}-(m_1+m_2)\bm{p}'_{3}]^2} {2(m_1+m_2+m_3)^2\hbar^2c^2}}.      \label{eq:RHj3hxp} 
\end{eqnarray} }%
The superscript `$'$' denotes the baryon coordinate or momentum in the rest frame of the $h_1h_2h_3$-cluster.
The width parameter $\sigma_1=\sqrt{\frac{m_3(m_1+m_2)(m_1+m_2+m_3)} {m_1m_2(m_1+m_2)+m_2m_3(m_2+m_3)+m_3m_1(m_3+m_1)}} R_{H_{j}}$,
and $\sigma_2=\sqrt{\frac{4m_1m_2(m_1+m_2+m_3)^2} {3(m_1+m_2)[m_1m_2(m_1+m_2)+m_2m_3(m_2+m_3)+m_3m_1(m_3+m_1)]}} R_{H_{j}}$.

Substituting Eqs.~(\ref{eq:RHj3h}) and (\ref{eq:RHj3hxp}) into Eq.~(\ref{eq:fHj3hgeneral1}), we have
{\setlength\arraycolsep{0.2pt}
\begin{eqnarray}
&& f_{H_{j}}(\bm{p})=8^2N_{h_1h_2h_3} g_{H_{j}} \int d\bm{x}_1d\bm{x}_2d\bm{x}_3 d\bm{p}_1 d\bm{p}_2d\bm{p}_3  e^{-\frac{(\bm{x}'_1-\bm{x}'_2)^2}{2\sigma_1^2}}   \nonumber   \\
&&~~~~~~ \times   e^{-\frac{2(\frac{m_1\bm{x}'_1}{m_1+m_2}+\frac{m_2\bm{x}'_2}{m_1+m_2}-\bm{x}'_3)^2}{3\sigma_2^2}}   f^{(n)}_{h_1h_2h_3}(\bm{x}_1,\bm{x}_2,\bm{x}_3;\bm{p}_1,\bm{p}_2,\bm{p}_3)     \nonumber  \\
&&~~~~~~ \times  e^{-\frac{2\sigma_1^2(m_2\bm{p}'_{1}-m_1\bm{p}'_{2})^2}{(m_1+m_2)^2\hbar^2c^2}}
 e^{-\frac{3\sigma_2^2[m_3\bm{p}'_{1}+m_3\bm{p}'_{2}-(m_1+m_2)\bm{p}'_{3}]^2} {2(m_1+m_2+m_3)^2\hbar^2c^2}}
	\delta(\displaystyle{\sum^3_{i=1}} \bm{p}_i-\bm{p}).      \label{eq:fHj3h}  
\end{eqnarray} }%

Approximating the gaussian form of the momentum-dependent kernel function to be $\delta$ function form and integrating $\bm{p}_1$, $\bm{p}_2$ and $\bm{p}_3$ from Eq.~(\ref{eq:fHj3h}), we can obtain
\begin{widetext}
{\setlength\arraycolsep{0.2pt}
\begin{eqnarray}
f_{H_{j}}(\bm{p})=&& 8^2 N_{h_1h_2h_3} g_{H_{j}} \int d\bm{x}_1d\bm{x}_2d\bm{x}_3d\bm{p}_1d\bm{p}_2d\bm{p}_3   f^{(n)}_{h_1h_2h_3}(\bm{x}_1,\bm{x}_2,\bm{x}_3;\bm{p}_1,\bm{p}_2,\bm{p}_3) 
   e^{-\frac{(\bm{x}'_1-\bm{x}'_2)^2}{2\sigma_1^2}} e^{-\frac{2(\frac{m_1\bm{x}'_1}{m_1+m_2}+\frac{m_2\bm{x}'_2}{m_1+m_2}-\bm{x}'_3)^2}{3\sigma_2^2}}  \nonumber   \\
		&&~~\times   (\frac{\hbar c\sqrt{\pi}}{\sqrt{2} \sigma_1})^3  (1+\frac{m_1}{m_2})^3
		\delta(\bm{p}'_{1}-\frac{m_1}{m_2} \bm{p}'_{2})   (\frac{\sqrt{2\pi} \hbar c}{\sqrt{3} \sigma_2})^3 (1+\frac{m_1}{m_3}+\frac{m_2}{m_3})^3
		\delta(\bm{p}'_{1}+\bm{p}'_{2}-\frac{m_1+m_2}{m_3} \bm{p}'_{3}) \delta(\displaystyle{\sum^2_{i=1}} \bm{p}_i-\bm{p})          \nonumber   \\	
=&&  8^2 N_{h_1h_2h_3} g_{H_{j}} \int d\bm{x}_1d\bm{x}_2d\bm{x}_3d\bm{p}_1d\bm{p}_2d\bm{p}_3  f^{(n)}_{h_1h_2h_3}(\bm{x}_1,\bm{x}_2,\bm{x}_3;\bm{p}_1,\bm{p}_2,\bm{p}_3)
    e^{-\frac{(\bm{x}'_1-\bm{x}'_2)^2}{2\sigma_1^2}}e^{-\frac{2(\frac{m_1\bm{x}'_1}{m_1+m_2}+\frac{m_2\bm{x}'_2}{m_1+m_2}-\bm{x}'_3)^2}{3\sigma_2^2}}  \nonumber   \\
		&&~~\times 	(\frac{\hbar c\sqrt{\pi}}{\sqrt{2} \sigma_1})^3  (1+\frac{m_1}{m_2})^3
    	\gamma	\delta(\bm{p}_{1}-\frac{m_1}{m_2} \bm{p}_{2})   (\frac{\sqrt{2\pi} \hbar c}{\sqrt{3} \sigma_2})^3 (1+\frac{m_1}{m_3}+\frac{m_2}{m_3})^3
		\gamma \delta(\bm{p}_{1}+\bm{p}_{2}-\frac{m_1+m_2}{m_3} \bm{p}_{3})  \delta(\displaystyle{\sum^2_{i=1}} \bm{p}_i-\bm{p})          \nonumber   \\             
=&& 8^2 N_{h_1h_2h_3} g_{H_{j}} \gamma{^2}  (\frac{\hbar^{2} c^{2}\pi}{\sqrt{3}\sigma_1\sigma_2})^3 
  \int d\bm{x}_1d\bm{x}_2d\bm{x}_3  f^{(n)}_{h_1h_2h_3}(\bm{x}_1,\bm{x}_2,\bm{x}_3;\frac{m_1 \bm{p}}{m_1+m_2+m_3} , \frac{m_2 \bm{p}}{m_1+m_2+m_3},\frac{m_3 \bm{p}}{m_1+m_2+m_3}) \nonumber   \\
  &&~~~~~~~~~~~~~~~~~~~~~~~~~~~~~~~~~~~~~~~~~~~~~~~~\times e^{-\frac{(\bm{x}'_1-\bm{x}'_2)^2}{2\sigma_1^2}} e^{-\frac{2(\frac{m_1\bm{x}'_1}{m_1+m_2}+\frac{m_2\bm{x}'_2}{m_1+m_2}-\bm{x}'_3)^2}{3\sigma_2^2}}. \label{eq:fH3-p}   
\end{eqnarray} }%
		
Changing coordinate variables in Eq.~(\ref{eq:fH3-p}) to be $\bm{Y}= (m_1\bm{x}_1+m_2\bm{x}_2+m_3\bm{x}_3)/(m_1+m_2+m_3)$, $\bm{r}_1= (\bm{x}_1-\bm{x}_2)/\sqrt{2}$ and 
$\bm{r}_2=\sqrt{\frac{2}{3}} (\frac{m_1\bm{x}_1}{m_1+m_2}+\frac{m_2\bm{x}_2}{m_1+m_2}-\bm{x}_3)$, we have  
{\setlength\arraycolsep{0pt}
\begin{eqnarray}
f_{H_j}(\bm{p})&=& 8^2 N_{h_1h_2h_3} g_{H_{j}} \gamma{^2}  (\frac{\hbar^{2} c^{2}\pi}{\sqrt{3}\sigma_1\sigma_2})^3
 3^{3/2} \int d\bm{Y}d\bm{r_1}d\bm{r_2}  f^{(n)}_{h_1h_2h_3}(\bm{Y},\bm{r}_1,\bm{r}_2;\frac{m_1 \bm{p}}{m_1+m_2+m_3}, \frac{m_2 \bm{p}}{m_1+m_2+m_3}, \frac{m_3 \bm{p}}{m_1+m_2+m_3})
 e^{-\frac{\bm{r}_1'^2}{\sigma_1^2}} e^{-\frac{\bm{r}_2'^2}{\sigma_2^2}}. ~~~~~~    \label{eq:fH3-Xr}  
\end{eqnarray} }%
We also assume the center of mass coordinate in joint distribution is factorized, i.e.,
{\setlength\arraycolsep{0pt}
\begin{eqnarray}
&&~~~ 3^{3/2} f^{(n)}_{h_1h_2h_3}(\bm{Y},\bm{r}_1,\bm{r}_2;\frac{m_1  \bm{p}}{m_1+m_2+m_3}, \frac{m_2  \bm{p}}{m_1+m_2+m_3},\frac{m_3 \bm{p}}{m_1+m_2+m_3} ) \nonumber   \\
&& =  f^{(n)}_{h_1h_2h_3}(\bm{Y}) f^{(n)}_{h_1h_2h_3}(\bm{r}_1,\bm{r}_2;\frac{m_1  \bm{p}}{m_1+m_2+m_3},\frac{m_2 \bm{p}}{m_1+m_2+m_3}, \frac{m_3  \bm{p}}{m_1+m_2+m_3}).  \label{eq:fh1h2h3-Xr}  
\end{eqnarray} }%
Substituting Eq.~(\ref{eq:fh1h2h3-Xr}) into Eq.~(\ref{eq:fH3-Xr}), we have
{\setlength\arraycolsep{0pt}
\begin{eqnarray}
&& f_{H_{j}}(\bm{p})= 8^2 N_{h_1h_2h_3} g_{H_{j}} \gamma{^2}  (\frac{\hbar^{2} c^{2}\pi}{\sqrt{3}\sigma_1\sigma_2})^3 
  \int d\bm{r_1}d\bm{r_2}  f^{(n)}_{h_1h_2h_3}(\bm{r}_1,\bm{r}_2;\frac{m_1  \bm{p}}{m_1+m_2+m_3}, \frac{m_2 \bm{p}}{m_1+m_2+m_3},  \frac{m_3  \bm{p}}{m_1+m_2+m_3}) 
    e^{-\frac{\bm{r}_1'^2}{\sigma_1^2}} e^{-\frac{\bm{r}_2'^2}{\sigma_2^2}}.   \label{eq:fH3-r} 
\end{eqnarray} }%
		
Adopting gaussian forms for the relative coordinate distributions~\cite{Mrowczynski:2016xqm,Wang:2020zaw}, we have
{\setlength\arraycolsep{0pt}
\begin{eqnarray}
&&~~~ f^{(n)}_{h_1h_2h_3}(\bm{r}_1,\bm{r}_2;\frac{m_1  \bm{p}}{m_1+m_2+m_3}, \frac{m_2 \bm{p}}{m_1+m_2+m_3},\frac{m_3 \bm{p}}{m_1+m_2+m_3}) \nonumber   \\ 
&&=  \frac{1}{[\pi C_1 R_f^2(\bm{p})]^{3/2}} e^{-\frac{\bm{r}_1^2}{C_1 R_f^2(\bm{p})}} \frac{1}{[\pi C_2 R_f^2(\bm{p})]^{3/2}}  
  e^{-\frac{\bm{r}_2^2}{C_2 R_f^2(\bm{p})}} f^{(n)}_{h_1h_2h_3}(\frac{m_1  \bm{p}}{m_1+m_2+m_3}, \frac{m_2 \bm{p}}{m_1+m_2+m_3},\frac{m_3 \bm{p}}{m_1+m_2+m_3}).  \label{eq:fh1h2h3-rp}    
\end{eqnarray} }%
Comparing relations of $\bm{r}_1$, $\bm{r}_2$ with $\bm{x}_1$, $\bm{x}_2$, $\bm{x}_3$ to that of $\bm{r}$ with $\bm{x}_1$, $\bm{x}_2$ in Sec.~\ref{2hcoHj}, 
we see that $C_1$ is equal to $C_w$ and $C_2$ is $4C_w/3$ when ignoring the mass difference of $m_1$ and $m_2$~\cite{Mrowczynski:2016xqm,Wang:2020zaw}. 
Substituting Eq.~(\ref{eq:fh1h2h3-rp}) into Eq.~(\ref{eq:fH3-r}) and considering the Lorentz transformation, we integrate from the relative coordinate variables and obtain 
{\setlength\arraycolsep{0pt}
\begin{eqnarray}
f_{H_{j}}(\bm{p}) &=& \frac{64 \pi^3 \hbar^6 c^6 g_{H_{j}} \gamma^2}{3\sqrt{3} \left[C_1 R_f^2(\bm{p})+\sigma_1^2\right] \sqrt{C_1 [R_f(\bm{p})/\gamma]^2+\sigma_1^2}
\left[C_2 R_f^2(\bm{p})+\sigma_2^2\right] \sqrt{C_2 [R_f(\bm{p})/\gamma]^2+\sigma_2^2}}    \nonumber   \\
		&& \times f_{h_1h_2h_3}(\frac{m_1  \bm{p}}{m_1+m_2+m_3},  \frac{m_2 \bm{p}}{m_1+m_2+m_3},\frac{m_3 \bm{p}}{m_1+m_2+m_3}) .   \label{eq:fHj3h-}
\end{eqnarray} }%
Ignoring correlations between $h_1$, $h_2$ and $h_3$, we have the three-dimensional momentum distribution of $H_j$ as
{\setlength\arraycolsep{0pt}
\begin{eqnarray}
f_{H_{j}}(\bm{p}) &=& \frac{64 \pi^3 \hbar^6 c^6 g_{H_{j}} \gamma^2}{3\sqrt{3}\left[C_1 R_f^2(\bm{p})+\sigma_1^2\right] \sqrt{C_1 [R_f(\bm{p})/\gamma]^2+\sigma_1^2}
\left[C_2 R_f^2(\bm{p})+\sigma_2^2\right] \sqrt{C_2 [R_f(\bm{p})/\gamma]^2+\sigma_2^2}}     \nonumber   \\
&& \times    f_{h_1}(\frac{m_1\bm{p}}{m_1+m_2+m_3})f_{h_2}(\frac{m_2\bm{p}}{m_1+m_2+m_3})  f_{h_3}(\frac{m_3\bm{p}}{m_1+m_2+m_3}) .     \label{eq:fHj3h-approx}
\end{eqnarray} }%
		
Finally we have the Lorentz invariant momentum distribution as
{\setlength\arraycolsep{0.2pt}
\begin{eqnarray}
f_{H_j}^{(inv)}(p_{T},y) &=& \frac{64 \pi^3 \hbar^6 c^6 g_{H_{j}} }{3\sqrt{3}\left[C_1 R_f^2(p_T,y)+\sigma_1^2\right] \sqrt{C_1 [R_f(p_T,y)/\gamma]^2+\sigma_1^2}
\left[C_2 R_f^2(p_T,y)+\sigma_2^2\right] \sqrt{C_2 [R_f(p_T,y)/\gamma]^2+\sigma_2^2}}   \frac{m_{H_j}}{m_1m_2m_3}   \nonumber   \\
&&  \times        f_{h_1}^{(inv)}(\frac{m_1p_{T}}{m_1+m_2+m_3},y)  f_{h_2}^{(inv)}(\frac{m_2p_{T}}{m_1+m_2+m_3},y) f_{h_3}^{(inv)}(\frac{m_3p_{T}}{m_1+m_2+m_3},y).    \label{eq:pt-Hj3h}
\end{eqnarray} }%
\end{widetext}

As a short summary of this section, we want to state that Eqs.~(\ref{eq:pt-Hj2h}) and (\ref{eq:pt-Hj3h}) give the relationships of dibaryon states and tribaryon states with primordial baryons in momentum space in the laboratory frame.
They clearly show effects of different factors on dibaryon or tribaryon production such as the whole hadronic system scale as well as the sizes of the formed composite objects.
They can be directly used to calculate the $p_T$ distributions, the rapidity distributions and even the total yields of light (hyper-)nuclei as long as the primordial baryon momentum distributions are given.
Formulas for the antiparticles are the same as these dibaryon and tribaryon states, and we leave out the duplication.
Their applications at midrapidity (i.e., $y=0$) in heavy ion collisions at the LHC will be shown in the following sections.

\section{Results of light nuclei}    \label{lightnuclei}

In this section, we use the coalescence model to study productions of $d$, $^3\overline{\text{He}}$ and $\bar t$ at midrapidity in Pb+Pb collisions at $\sqrt{s_{NN}}=5.02$ TeV.
We first calculate the coalescence factors $B_2$, $B_3$ and discuss their centrality and $p_T$-dependent behaviors.
We then compute the $p_T$ spectra of $d$, $^3\overline{\text{He}}$ and $\bar t$.
We finally calculate the averaged transverse momenta $\langle p_T \rangle$, the yield rapidity densities $dN/dy$ and yield ratios of different light nuclei.

\subsection{The coalescence factor of light nuclei} 
The coalescence factor $B_A$ is defined as
\begin{eqnarray}
 B_A (p_T) = f_{d,\mathrm{^{3}He},t}^{(inv)}(p_T) /  \left\{    \left[ f_{p}^{(inv)}(\frac{p_T}{A}) \right]^Z  \left[ f_{n}^{(inv)}(\frac{p_T}{A}) \right]^{A-Z}   \right\}, ~~ \label{eq:BA}
\end{eqnarray}
where $A$ is the mass number and $Z$ is the charge of the light nuclei.
From Eqs.~(\ref{eq:pt-Hj2h}) and (\ref{eq:pt-Hj3h}), we respectively have for $d$, $^3$He and $t$
{\setlength\arraycolsep{0.2pt}
\begin{eqnarray}
&&  B_2(p_{T}) = \frac{ m_dg_d (\sqrt{2\pi}\hbar c)^3 }{m_pm_n \left[ C_w R_f^2({p_{T}})+\sigma_d^2 \right] \sqrt{C_w [\frac{R_f(p_T)}{\gamma}]^2+\sigma_d^2}},~~ \label{eq:B2-d}          \\
&&  B_3(p_{T}) =  \frac{64 \pi^3 \hbar^6 c^6 g_{\mathrm{^{3}He}} }{3\sqrt{3}\left[C_1 R_f^2(p_{T})+\sigma_{\mathrm{^{3}He}}^2 \right] \sqrt{C_1 [R_f(p_{T})/\gamma]^2+\sigma_{\mathrm{^{3}He}}^2} }  \nonumber \\
&&~~~\times \frac{m_{\mathrm{^{3}He}}}{m^2_pm_n\left[C_2 R_f^2(p_{T})+\sigma_{\mathrm{^{3}He}}^2\right]  \sqrt{C_2 [R_f(p_{T})/\gamma]^2+\sigma_{\mathrm{^{3}He}}^2}},~~ \label{eq:B3-He3}  \\
&&  B_3(p_{T}) = \frac{64 \pi^3 \hbar^6 c^6 g_t }{3\sqrt{3}\left[C_1 R_f^2(p_{T})+\sigma_t^2\right] \sqrt{C_1 [R_f(p_{T})/\gamma]^2+\sigma_t^2}}    \nonumber \\
&&~~~\times \frac{m_t}{m_pm^2_n\left[C_2 R_f^2(p_{T})+\sigma_t^2\right] \sqrt{C_2 [R_f(p_{T})/\gamma]^2+\sigma_t^2}}.  \label{eq:B3-t}
\end{eqnarray} }%
Here $\sigma_d=\sqrt{\frac{4}{3}} R_d$, and the root-mean-square radius of the deuteron $R_d=2.1421$ fm~\cite{Angeli:2013epw}.
$\sigma_{\mathrm{^{3}He}}=R_{\mathrm{^{3}He}}=1.9661$ fm and $\sigma_t=R_t=1.7591$ fm~\cite{Angeli:2013epw}.
$m_{p,n}$ denotes the nucleon mass and $m_{d,\mathrm{^{3}He},t}$ the mass of the $d$, $^3$He and $t$.
Eqs.~(\ref{eq:B2-d}-\ref{eq:B3-t}) show that $p_T$-dependent behaviors of $B_2$ and $B_3$ are related with the Lorentz contraction factor $\gamma$ and $R_f(p_{T})$.

\begin{figure}[htbp]%
	\centering
	\includegraphics[width=0.8\hsize]{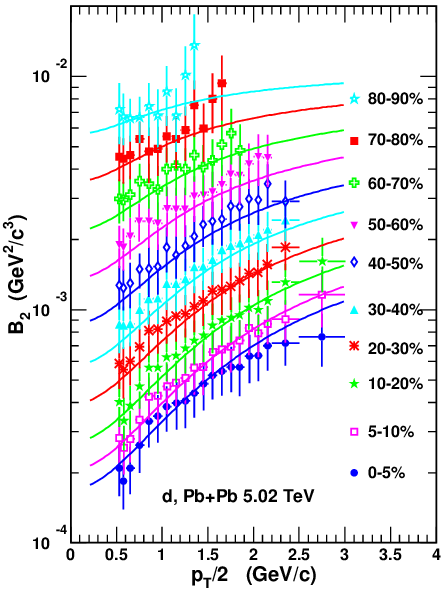}
	\caption{The $B_2$ of $d$ as a function of $p_T/2$ in different centralities in Pb+Pb collisions at $\sqrt{s_{NN}}=5.02$ TeV.
		Symbols with error bars are experimental data~\cite{Puccio:2017zir} and solid lines are theoretical results.}
	\label{fig:B2}
\end{figure}

\begin{figure*}[htbp]%
	\centering
	\includegraphics[width=0.8\textwidth]{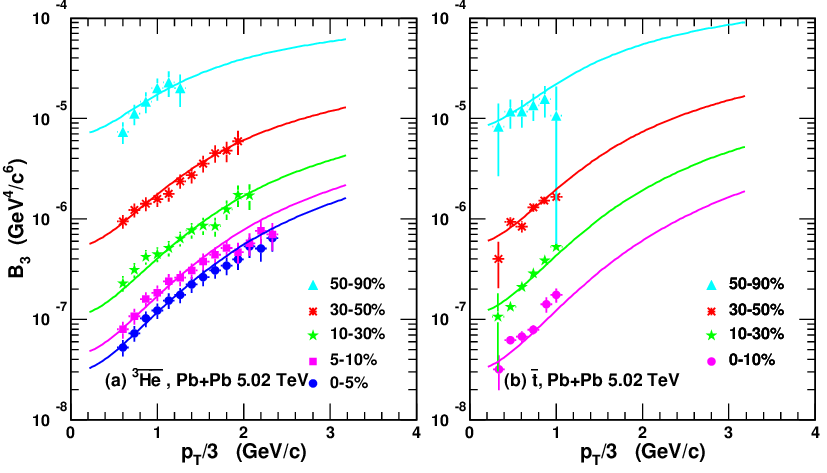}
	\caption{The $B_3$ of (a) $^3\overline{\text{He}}$ and (b) $\bar t$ as a function of $p_T/3$ in different centralities in Pb+Pb collisions at $\sqrt{s_{NN}}=5.02$ TeV.
		Symbols with error bars are experimental data~\cite{ALICE:2022veq} and solid lines are theoretical results.}
	\label{fig:B3}
\end{figure*}

To further compute $B_2$ and $B_3$, the specific form of $R_{f}(p_{T})$ is necessary.
Similar as in Ref.~\cite{Wang:2023rpd}, the dependence of $R_{f}(p_{T})$ on centrality and $p_T$ is considered to factorize into a linear dependence on the cube root of the pseudorapidity density of charged particles $(dN_{ch}/d\eta)^{1/3}$ and a power-law dependence on the transverse mass of the formed light nucleus $m_T$~\cite{ALICE:2015tra}.
So we have
{\setlength\arraycolsep{0.2pt}
\begin{eqnarray}
	R_{f}(p_{T})=a \times (dN_{ch}/d\eta)^{1/3} \times \left(\sqrt{p_T^2+m_{d,\mathrm{^{3}He},t}^2} \right)^b, \label{Eq:Rf}
\end{eqnarray} }%
where $a$ and $b$ are free parameters.
Their values in Pb+Pb collisions at $\sqrt{s_{NN}}=5.02$ TeV are (0.70,-0.31) for $d$ and (0.66,-0.31) for $^3$He and $t$, which are determined by reproducing the data of the $p_T$ spectra of $d$ and $^3$He in the most central 0-5\% centrality.
Here $b$ is set to be centrality independent, which is consistent with that in hydrodynamics~\cite{Chakraborty:2020tym} and that in STAR measurements of two-pion interferometry in central and simicentral Au+Au collisions~\cite{STAR:2004qya}.
Note that $b<0$ gives decreasing $R_f$ with the increasing $p_T$.
This means that nucleons with stronger collective motions have larger probability to emit from the same spatial location.
Such coordinate-momentum correlation is naturally given by the gaussian form of the relative coordinate distribution in Eqs.~(\ref{eq:fh1h2-rp}) and (\ref{eq:fh1h2h3-rp}).
$a$ is also set to be centrality-independent, the same as that in our previous work~\cite{Wang:2023rpd}.

We use the data of $dN_{ch}/d\eta$ in Ref.~\cite{ALICE:2019hno} to evaluate $R_f$, and then compute coalescence factors $B_2$ and $B_3$.
Fig.~\ref{fig:B2} shows $B_2$ of $d$ as a function of the transverse momentum scaled by the mass number $p_T/2$ in different centralities in Pb+Pb collisions at $\sqrt{s_{NN}}=5.02$ TeV.
Symbols with error bars are experimental data~\cite{Puccio:2017zir} and solid lines are theoretical results of the coalescence model.
From Fig.\ref{fig:B2}, one can see from central to peripheral collisions, $B_2$ exhibits an increasing trend, which is due to the decreasing size of the created hadronic system.
For the same centrality, $B_2$ increase as a function of $p_T/2$.
This increase behavior results on one hand from the Lorentz contraction factor $\gamma$, which has been studied in Ref.~\cite{Wang:2020zaw}.
On the other hand, it results from the decreasing $R_f$ with increasing $p_T$.
The rising behavior of the experimental data as a function of $p_T/2$ from central to peripheral collisions can be quantitively described by the coalescence model. 

Fig.~\ref{fig:B3} shows $B_3$ of $^3\overline{\text{He}}$ and that of $\bar t$ as a function of $p_T/3$ in different centralities in Pb+Pb collisions at $\sqrt{s_{NN}}=5.02$ TeV.
Symbols with error bars are experimental data~\cite{ALICE:2022veq} and solid lines are theoretical results.
Similarly as $B_2$, experimental data of $B_3$ also exhibits a rising trend as a function of $p_T/3$, which is reproduced well by the coalescence model from central to peripheral collisions.
Fig.~\ref{fig:B2} and Fig.~\ref{fig:B3} show that the centrality and momentum dependent behaviors of $B_2$ and $B_3$ in Pb+Pb collisions at $\sqrt{s_{NN}}=5.02$ TeV are simultaneously explained by the coalescence model.
The influencing factors of $B_2$ and $B_3$ are explicitly unfolded, as shown in Eqs.~(\ref{eq:B2-d}-\ref{eq:B3-t}).
Some other models based on transport approach are also used to study behaviors of $B_A$ in heavy ion collisions at the high LHC energies~\cite{Oliinychenko:2018ugs,Oliinychenko:2018odl,Bailung:2023dpv,Liu:2022vbg}.
All the results from these different models can help cross understand production properties of light nuclei from different aspects.

\subsection{The $p_T$ spectra of light nuclei} 

\begin{figure}[!htb]
\includegraphics[width=0.8\hsize]{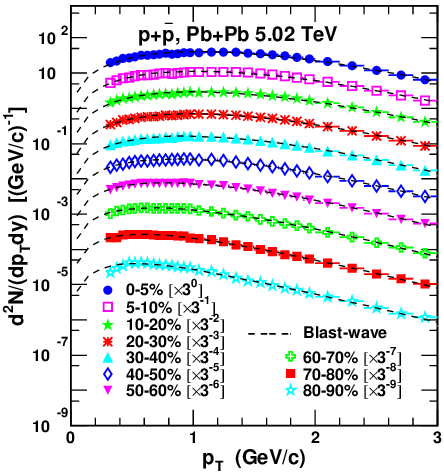}
\caption{The $p_T$ spectra of prompt protons plus antiprotons in different centralities in Pb+Pb collisions at $\sqrt{s_{NN}}=5.02$  TeV.
Symbols with error bars are experimental data~\cite{ALICE:2019hno}, and dashed lines are results of the blast-wave model.}
\label{fig:ppt}
\end{figure}

The $p_T$ spectra of primordial nucleons are necessary inputs for computing $p_T$ distributions of light nuclei in the coalescence model.
We here use the blast-wave model to get $p_T$ distribution functions of primordial protons by fitting the experimental data of prompt (anti)protons in Ref.~\cite{ALICE:2019hno}.
The blast-wave function~\cite{Schnedermann:1993ws} is given as
{\setlength\arraycolsep{0.2pt}
\begin{eqnarray}
\frac{d^{2}N}{2\pi p_{T}dp_{T}dy}  \propto  &&  \int_{0}^{R} r dr m_T I_0 \left(\frac{p_Tsinh\rho}{T_{kin}}\right)   K_1\left(\frac{m_Tcosh\rho}{T_{kin}}\right), ~~~~~~  \label{eq:BWfitfunc}
\end{eqnarray} }%
where $r$ is the radial distance in the transverse plane and $R$ is the radius of the fireball.
$I_0$ and $K_1$ are the modified Bessel functions, and the velocity profile $\rho=tanh^{-1}[\beta_s(\frac{r}{R})^n]$.
The surface velocity $\beta_s$, the kinetic freeze-out temperature $T_{kin}$ and $n$ are fitting parameters.

Fig.~\ref{fig:ppt} shows the $p_T$ spectra of prompt protons plus antiprotons in different centralities in Pb+Pb collisions at $\sqrt{s_{NN}}= 5.02$ TeV.
Symbols with error bars are experimental data~\cite{ALICE:2019hno}, and dashed lines are the results of the blast-wave model.
The $p_T$ spectra in different centralities are scaled by different factors for clarity as shown in the figure.
For the primordial neutron $p_T$ spectra, we adopt the same as those of primordial protons as we focus on light nucleus production at midrapidity at so high LHC energy that the isospin symmetry is well satisfied.

\begin{figure}[!htb]
\includegraphics[width=0.8\hsize]{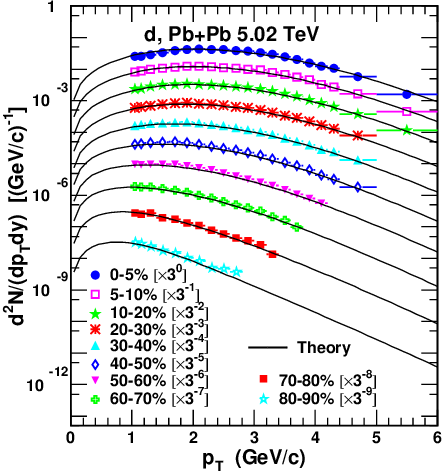}
\caption{The $p_T$ spectra of deuterons in different centralities in Pb+Pb collisions at $\sqrt{s_{NN}}=5.02$ TeV. 
Symbols with error bars are experimental data~\cite{Puccio:2017zir} and solid lines are theoretical results.}
\label{fig:dpt}
\end{figure}

\begin{figure*}[!htb]
\includegraphics[width=0.8\hsize]{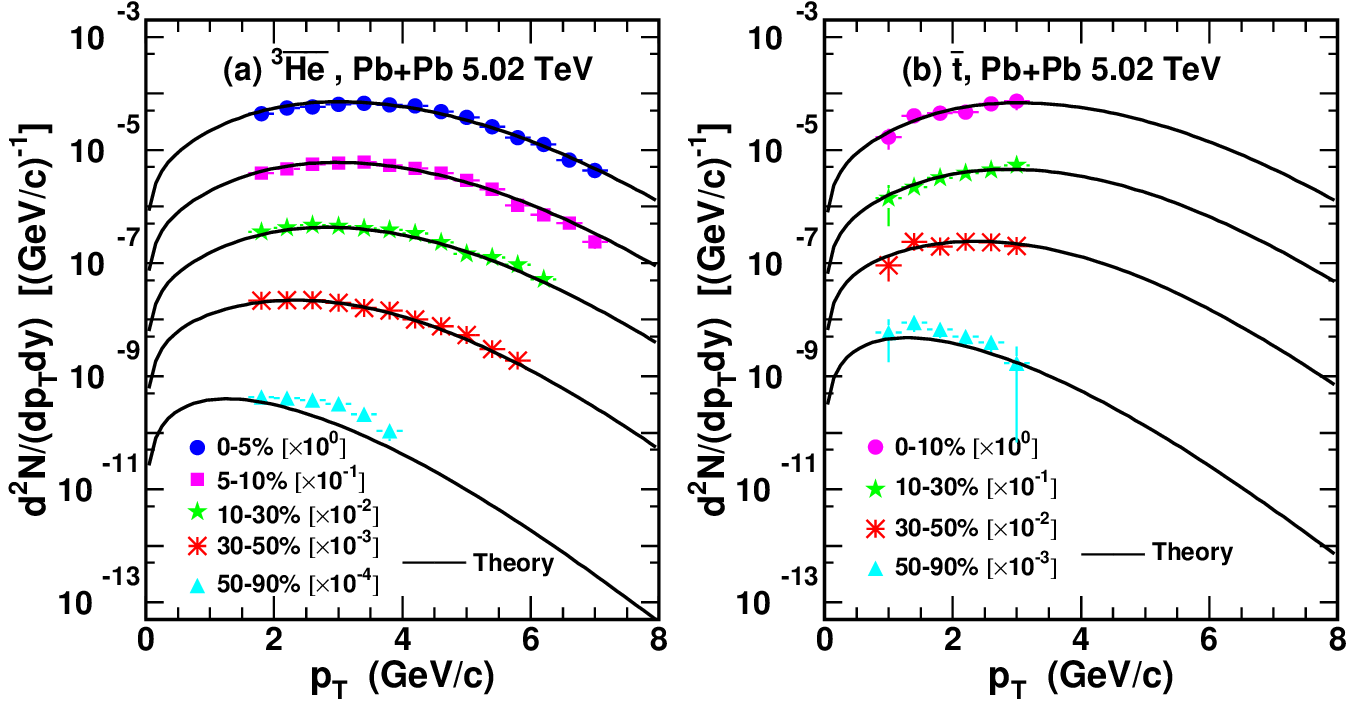}
\caption{The $p_T$ spectra of (a) $^3\overline{\text{He}}$ and (b) $\bar t$ in different centralities in Pb+Pb collisions at $\sqrt{s_{NN}}=5.02$ TeV. 
Filled symbols with error bars are experimental data~\cite{ALICE:2022veq} and solid lines are theoretical results.}
\label{fig:He3tpt}
\end{figure*}

We first calculate the $p_T$ spectra of deuterons in Pb+Pb collisions at $\sqrt{s_{NN}}=5.02$ TeV in $0-5$\%, $5-10$\%, $10-20$\%, $20-30$\%, $30-40$\%, $40-50$\%, $50-60$\%,  $60-70$\%, $70-80$\% and $80-90$\% centralities.
Different solid lines scaled by different factors for clarity in Fig.~\ref{fig:dpt} are our theoretical results.
Symbols with error bars are experimental data from the ALICE collaboration~\cite{ALICE:2022veq}. 
We then compute the $p_T$ spectra of $^3\overline{\text{He}}$ and $\bar t$ in Pb+Pb collisions at $\sqrt{s_{NN}}=5.02$ TeV in different centralities.
Different solid lines in Fig.~\ref{fig:He3tpt} are our theoretical results, which agree with the available data denoted by filled symbols~\cite{ALICE:2022veq}.
From Fig.~\ref{fig:dpt} and Fig.~\ref{fig:He3tpt}, one can see the nucleon coalescence is still the dominant mechanism for light nuclei production in Pb+Pb collisions at $\sqrt{s_{NN}}=5.02$ TeV.
More precise measurements for $^3\overline{\text{He}}$ and $\bar t$ in wide $p_T$ range in the forthcoming future can help further test the coalescence mechanism, especially in peripheral Pb+Pb collisions.

\subsection{Averaged transverse momenta and yield rapidity densities of light nuclei} 

\begin{table*}[!htb]
\begin{center}
\caption{Averaged transverse momenta $\langle p_T\rangle$ and yield rapidity densities $dN/dy$ of $d$, $^3\overline{\text{He}}$ and $\bar t$ in different centralities in Pb+Pb collisions at $\sqrt{s_{NN}}=5.02$ TeV. Experimental data in the third and fifth columns are from Refs.~\cite{Puccio:2017zir,ALICE:2022veq}. Theoretical results are in the fourth and sixth columns.}  \label{table:dNdypt-dHe3t}
\begin{tabular}{@{\extracolsep{\fill}}cccccccc@{\extracolsep{\fill}}}
\toprule
     &\multirow{2}{*}{Centrality}     &\multicolumn{2}{@{}c@{}}{$\langle p_T\rangle$ (GeV)}          &          &\multicolumn{2}{@{}c@{}}{$dN/dy$} \\
\cline{3-4}  \cline{6-7} 
                          &     &Data                                 &Theory                   &          &Data                                                       &Theory      \\
\hline
\multirow{10}{*}{$d$}
 &0-5\%                     &$2.45\pm0.00\pm0.09$    &$2.37$                  &          &$(1.19\pm0.00\pm0.21)\times10^{-1}$  &$1.22\times10^{-1}$    \\
 &5-10\%                   &$2.41\pm0.01\pm0.10$    &$2.33$                  &          &$(1.04\pm0.00\pm0.19)\times10^{-1}$  &$1.01\times10^{-1}$    \\
 &10-20\%                 &$2.34\pm0.00\pm0.11$    &$2.28$                  &          &$(8.42\pm0.02\pm1.50)\times10^{-2}$  &$7.86\times10^{-2}$    \\
 &20-30\%                 &$2.21\pm0.00\pm0.12$    &$2.18$                  &          &$(6.16\pm0.02\pm1.10)\times10^{-2}$  &$5.58\times10^{-2}$    \\
 &30-40\%                 &$2.05\pm0.00\pm0.12$    &$2.04$                  &          &$(4.25\pm0.01\pm0.75)\times10^{-2}$  &$3.82\times10^{-2}$    \\
 &40-50\%                 &$1.88\pm0.01\pm0.12$    &$1.87$                  &          &$(2.73\pm0.01\pm0.48)\times10^{-2}$  &$2.46\times10^{-2}$    \\
 &50-60\%                 &$1.70\pm0.01\pm0.11$    &$1.66$                  &          &$(1.62\pm0.01\pm0.28)\times10^{-2}$  &$1.47\times10^{-2}$    \\
 &60-70\%                 &$1.46\pm0.01\pm0.12$    &$1.45$                  &          &$(8.35\pm0.14\pm1.43)\times10^{-3}$  &$7.58\times10^{-3}$    \\
 &70-80\%                 &$1.27\pm0.02\pm0.11$    &$1.25$                  &          &$(3.52\pm0.06\pm0.63)\times10^{-3}$  &$3.22\times10^{-3}$    \\
 &80-90\%                 &$1.09\pm0.02\pm0.40$    &$1.10$                  &          &$(1.13\pm0.03\pm0.23)\times10^{-3}$  &$0.925\times10^{-3}$    \\
\hline
\multirow{5}{*}{$^3\overline{\text{He}}$}
 &0-5\%                     &$3.465\pm0.013\pm0.154\pm0.144$    &$3.26$                  &          &$(24.70\pm0.28\pm2.29\pm0.30)\times10^{-5}$  &$25.6\times10^{-5}$    \\
 &5-10\%                   &$3.368\pm0.014\pm0.141\pm0.132$    &$3.21$                  &          &$(20.87\pm0.26\pm1.95\pm0.43)\times10^{-5}$  &$21.4\times10^{-5}$    \\
 &10-30\%                 &$3.237\pm0.021\pm0.157\pm0.150$    &$3.08$                  &          &$(15.94\pm0.31\pm1.53\pm0.34)\times10^{-5}$  &$14.8\times10^{-5}$    \\
 &30-50\%                 &$2.658\pm0.016\pm0.084\pm0.049$    &$2.64$                  &          &$(7.56\pm0.13\pm0.70\pm0.10)\times10^{-5}$  &$7.16\times10^{-5}$    \\
 &50-90\%                 &$2.057\pm0.023\pm0.090\pm0.027$    &$1.77$                  &          &$(1.19\pm0.08\pm0.16\pm0.14)\times10^{-5}$  &$0.931\times10^{-5}$    \\
\hline
\multirow{4}{*}{$\bar t$}
 &0-10\%                   &$3.368\pm0.241\pm0.060$    &$3.27$                  &          &$(24.45\pm1.75\pm2.71)\times10^{-5}$  &$24.6\times10^{-5}$    \\
 &10-30\%                 &$3.015\pm0.286\pm0.040$    &$3.11$                  &          &$(14.19\pm1.35\pm1.29)\times10^{-5}$  &$15.9\times10^{-5}$    \\
 &30-50\%                 &$2.524\pm0.593\pm0.180$    &$2.68$                  &          &$(7.24\pm1.70\pm0.65)\times10^{-5}$  &$7.97\times10^{-5}$    \\
 &50-90\%                 &$1.636\pm0.226\pm0.040$    &$1.80$                  &          &$(1.66\pm0.23\pm0.16)\times10^{-5}$  &$1.14\times10^{-5}$    \\
\botrule
\end{tabular}
\end{center}
\end{table*}

We here study the averaged transverse momenta $\langle p_T\rangle$ and yield rapidity densities $dN/dy$ of $d$, $^3\overline{\text{He}}$ and $\bar t$.
Our theoretical results are put in the fourth and sixth columns in Table \ref{table:dNdypt-dHe3t}.
Experimental data in the third and fifth columns are from Refs.~\cite{Puccio:2017zir,ALICE:2022veq}.
A clear decreasing trend for both $\langle p_T\rangle$ and $dN/dy$ from central to peripheral collisions is observed.
This is due to that in more central collisions more energy is deposited in the midrapidity region and collective evolution exists longer.
Theoretical results for $d$, $^3\overline{\text{He}}$ and $\bar t$ are consistent with the corresponding data within the experimental uncertainties except a very little underestimation for the $dN/dy$ of $\bar t$ in peripheral 50-90\% collision.
Such underestimation needs to be confirmed by future precise data.

\subsection{Yield ratios of light nuclei}

\begin{figure*}[!htb]
\includegraphics[width=0.72\hsize]{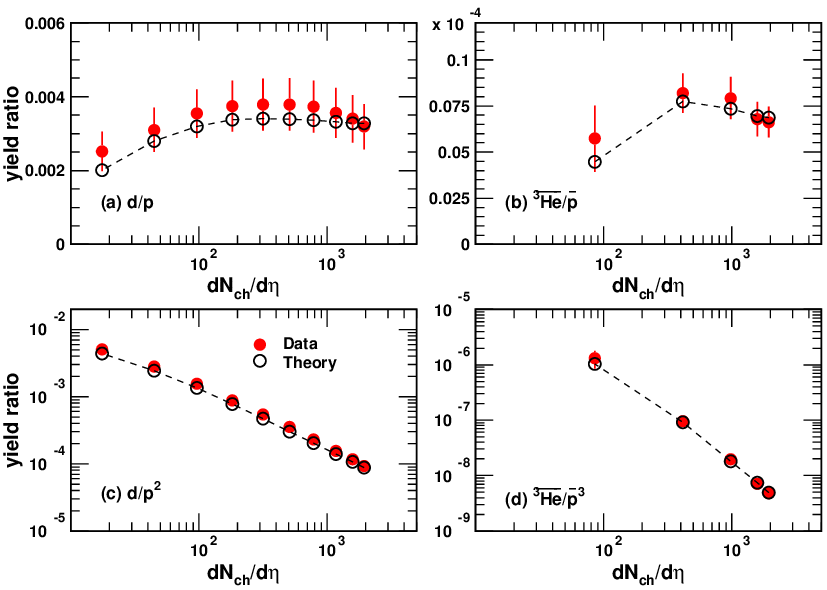}
\caption{Yield ratios (a) $d/p$, (b) $^3\overline{\text{He}}/\bar p$, (c) $d/p^2$ and (d) $^3\overline{\text{He}}/\bar p^3$ as a function of $dN_{ch}/d\eta$ in Pb+Pb collisions at $\sqrt{s_{NN}}=5.02$ TeV. 
Filled circles with error bars are experimental data~\cite{Puccio:2017zir,ALICE:2022veq} and open circles connected with dashed lines to guide the eye are theoretical results.}
\label{fig:Rdp}
\end{figure*}

\begin{figure*}[!htb]
\includegraphics[width=0.7\hsize]{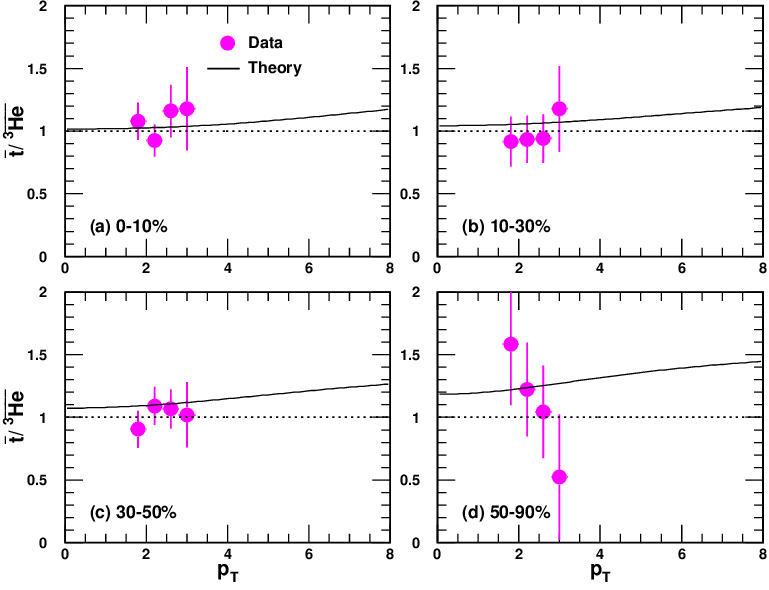}
\caption{Yield ratio $\bar t/^3\overline{\text{He}}$ as a function of $p_T$ in different centralities in Pb+Pb collisions at $\sqrt{s_{NN}}=5.02$ TeV.
Filled circles with error bars are experimental data~\cite{ALICE:2022veq} and solid lines are theoretical results.}
\label{fig:RtHe3pt}
\end{figure*}

\begin{figure}[!htb]
\includegraphics[width=0.85\hsize]{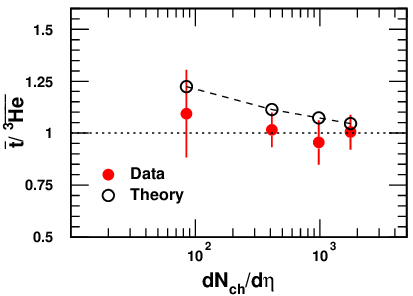}
\caption{Yield ratio $\bar t/^3\overline{\text{He}}$ as a function of $dN_{ch}/d\eta$ in Pb+Pb collisions at $\sqrt{s_{NN}}=5.02$ TeV.
Filled circles with error bars are experimental data~\cite{ALICE:2022veq} and open circles connected with dashed lines to guide the eye are theoretical results.}
\label{fig:RtHe3}
\end{figure}

Yield ratios of light nuclei carry information of intrinsic production correlations of different light nuclei and are predicted to have nontrivial behaviors~\cite{Wang:2023rpd}.
In this subsection we study centrality dependence of different yield ratios, such as $d/p$, $^3\overline{\text{He}}/\bar p$, $d/p^2$, $^3\overline{\text{He}}/\bar p^3$ and $\bar t/^3\overline{\text{He}}$.

Fig.~\ref{fig:Rdp} (a) and (b) show the $dN_{ch}/d\eta$ dependence of $d/p$ and $^3\overline{\text{He}}/\bar p$ in Pb+Pb collisions at $\sqrt{s_{NN}}=5.02$ TeV. 
Filled circles with error bars are experimental data~\cite{ALICE:2015wav}, and open circles connected with dashed lines to guide the eye are theoretical results.
From Eq.~(\ref{eq:pt-Hj2h}) we approximately have the $p_T$-integrated yield ratio
{\setlength\arraycolsep{0.2pt}
\begin{eqnarray}
         \frac{d}{p}
\propto && \frac{ N_p }{\langle R_f \rangle^3 \left(C_w +\frac{\sigma_{d}^2}{\langle R_f \rangle^2}\right)  \sqrt{\frac{C_w}{\langle\gamma\rangle^2}+\frac{\sigma_{d}^2}{\langle R_f \rangle^2}}} \nonumber \\ 
          =&& \frac{N_p}{\langle R_f \rangle^3/\langle\gamma\rangle}  \times \frac{ 1 }{ \left(C_w +\frac{\sigma_{d}^2}{\langle R_f \rangle^2}\right)  \sqrt{C_w+\frac{\sigma_{d}^2}{\langle R_f \rangle^2/\langle\gamma\rangle^2}}},   \label{eq:yRdp}  
\end{eqnarray} }%
where angle brackets denote the averaged values.
Eq.~(\ref{eq:yRdp}) gives that the $dN_{ch}/d\eta$-dependent behavior of $d/p$ is determined by two factors.
One is the nucleon number density $\frac{N_p}{\langle R_f \rangle^3/\langle\gamma\rangle}$ and the other is the suppression effect from the relative size of the formed light nuclei to the hadronic source system $\frac{\sigma_{d}}{\langle R_f \rangle}$.
Similar case holds for $^3\overline{\text{He}}/\bar p$.
The nucleon number density decreases especially from semicentral to central collisions~\cite{ALICE:2019hno}, which makes $d/p$ and $^3\overline{\text{He}}/\bar p$ decrease with the increasing $dN_{ch}/d\eta$.
The relative size $\frac{\sigma_{d}}{\langle R_f \rangle}$ decreases and its suppression effect becomes weak in large hadronic systems, which makes $d/p$ and $^3\overline{\text{He}}/\bar p$ increase with the increasing $dN_{ch}/d\eta$~\cite{Sun:2018mqq}.
For very high $dN_{ch}/d\eta$ area, difference of the suppression extents in different centralities becomes insignificant and the decreasing nucleon number density dominates the decreasing behavior of $d/p$ and $^3\overline{\text{He}}/\bar p$.
For low $dN_{ch}/d\eta$ area, different suppression extents of the relative size in different centralities make $d/p$ and $^3\overline{\text{He}}/\bar p$ increase as a function of $dN_{ch}/d\eta$.
The finally conjunct result from the nucleon number density and the suppression effect makes $d/p$ and $^3\overline{\text{He}}/\bar p$ first increase from peripheral to semicentral collisions and then decrease from semicentral to central collisions, just as shown from semicentral to central collisions of Fig.~\ref{fig:Rdp} (a) and (b).

Fig.~\ref{fig:Rdp} (c) and (d) show $d/p^{2}$ and $^3\overline{\text{He}}/\bar p^3$ as a function of $dN_{ch}/d\eta$ in Pb+Pb collisions at $\sqrt{s_{NN}}=5.02$ TeV. 
Filled circles with error bars are experimental data~\cite{ALICE:2015wav}.
Open circles connected with dashed lines to guide the eye are the theoretical results.
Both of them give explicit decreasing trend with the increasing $dN_{ch}/d\eta$, which are very different from the previous $d/p$ and $^3\overline{\text{He}}/\bar p$. 
Recalling that $d/p^{2}$ and $^3\overline{\text{He}}/\bar p^3$ represent the probability of any $pn$-pair coalescing into a deuteron and that of any $\bar p\bar p\bar n$-cluster coalescing into a $^3\overline{\text{He}}$.
This means that it is more difficult for any $pn$-pair or $\bar p\bar p\bar n$-cluster to recombine into a deuteron or $^3\overline{\text{He}}$ in larger hadronic system produced in more central collisions.

The yield ratio $t/^3$He is proposed to be a valuable probe to distinguish the thermal production and the coalescence production for light nuclei~\cite{Wang:2023rpd}.
In the coalescence picture, it is always larger than one and approaches to one at large $R_f$ where the suppression effect from the nucleus size can be ignored.
The smaller $R_f $, the higher deviation of ${t}/{^3\text{He}}$ from one.
The same case holds for $\bar t/^3\overline{\text{He}}$.
Fig.~\ref{fig:RtHe3pt} shows $\bar t/^3\overline{\text{He}}$ as a function of $p_T$ in Pb+Pb collisions at $\sqrt{s_{NN}}=5.02$ TeV in different centralities 0-10\%, 10-30\%, 30-50\% and 50-90\%.
Filled circles with error bars are experimental data~\cite{ALICE:2022veq} and solid lines are the theoretical results.
Reference line of one is plotted with dotted lines.
With the increasing $p_T$, $R_f$ decreases, so our theoretical results increase.
This feature is very different from that in the thermal model, where the expectation for this ratio is one~\cite{Andronic:2010qu}.
The trend of the data in 0-10\%, 10-30\% and 30-50\% centralities indicates an increasing hint but a final conclusion is hard to make due to the limited $p_T$ range and the large error bars.
Data in the peripheral 50-90\% centrality seems decrease, but further more precise measurements are needed to confirm.
More precise data in near future can be used to further distinguish production mechanisms of $^3\overline{\text{He}}$ and $\bar t$.

The $p_T$-integrated yield ratio $\bar t/^3\overline{\text{He}}$ as a function of $dN_{ch}/d\eta$ is in Fig.~\ref{fig:RtHe3}.
Filled circles with error bars are experimental data~\cite{ALICE:2022veq} and open circles connected with the dashed line to guide the eye are theoretical results.
Reference line of one is also plotted with the dotted line.
$\bar t/^3\overline{\text{He}}$ exhibits a decreasing trend.
This is because larger $dN_{ch}/d\eta$, i.e., larger $R_f$, makes $\bar t/^3\overline{\text{He}}$ decrease closer to one.
Theoretical results of $\bar t/^3\overline{\text{He}}$ in the coalescence model give non-flat behaviors as a function of $dN_{ch}/d\eta$.
This is due to different relative production suppression between $^3\overline{\text{He}}$ and $\bar t$ at different hadronic system scales.

\section{Results of the hypertriton and $\Omega$-hypernuclei}    \label{hynucleiResults}

In this section, we use the coalescence model in Sec.~\ref{model} to study productions of the hypertriton and $\Omega$-hypernuclei. 
We give the results of the $p_T$ spectra, the averaged $p_T$ and yield rapidity densities of the $^3_{\Lambda}$H.
We present the predictions of different $\Omega$-hypernuclei, such as $H(p\Omega^-)$, $H(n\Omega^-)$ and $H(pn\Omega^-)$.
We propose two groups of observables, both of which exhibit novel behaviors.
One group of observables are the averaged $p_T$ ratios of the light (hyper-)nuclei to the protons (hyperons), and the other are centrality-dependent yield ratios of theirs.

\subsection{The $p_T$ spectra of $\Lambda$ and $\Omega^-$ hyperons} 

\begin{figure}[!htb]
\includegraphics[width=0.83\hsize]{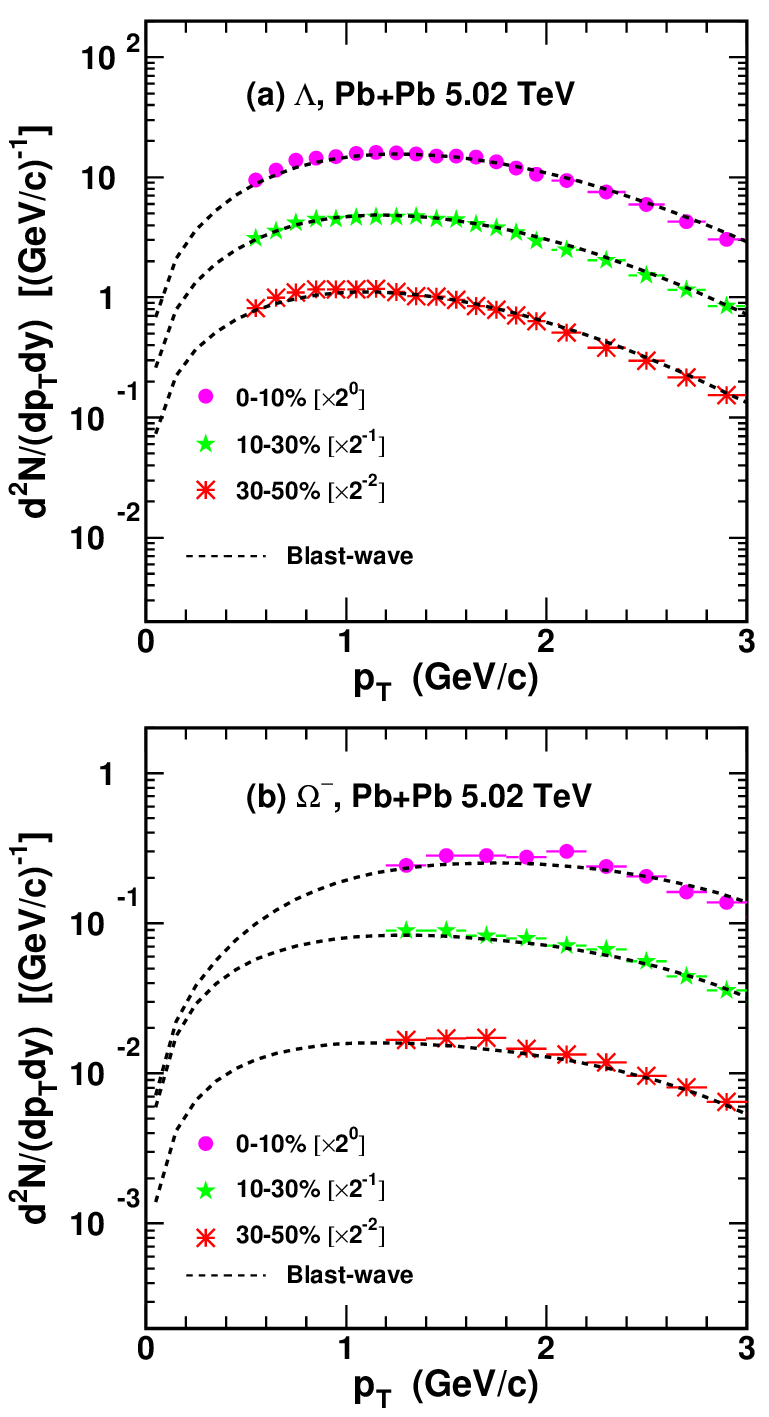}
\caption{The $p_T$ spectra of (a) $\Lambda$ and (b) $\Omega^-$ in Pb+Pb collisions at $\sqrt{s_{NN}}=5.02$ TeV. 
Symbols with error bars are experimental data~\cite{Kalinak:2017xll} and dashed lines are results of the blast-wave model.}
\label{fig:LamOmept}
\end{figure}

The $p_T$ spectra of $\Lambda$ and $\Omega^-$ hyperons are necessary for computing $p_T$ distributions of $^3_{\Lambda}$H and $\Omega$-hypernuclei.
We use the blast-wave model to get $p_T$ distribution functions by fitting the experimental data of $\Lambda$ and $\Omega^-$ in Pb+Pb collisions at $\sqrt{s_{NN}}=5.02$ TeV~\cite{Kalinak:2017xll}.
They are shown in Fig.~\ref{fig:LamOmept}.
Filled symbols with error bars are experimental data~\cite{Kalinak:2017xll}, and dashed lines are the results of the blast-wave model.
The $p_T$ spectra in 0-10\%, 10-30\% and 30-50\% centralities are scaled by $2^0$, $2^{-1}$ and $2^{-2}$, respectively, for clarity in the figure.
We have also studied the $p_T$ spectra of $\Lambda$ and $\Omega^-$ hyperons with the Quark Combination Model developed by the Shandong group (SDQCM) in another work~\cite{Chang:2023zbe}, where the results are consistent with the blast-wave model at low and intermediate $p_T$ regions. 
We in the following use these $\Lambda$ and $\Omega^-$ hyperons in Fig.~\ref{fig:LamOmept} to compute the productions of the $^3_{\Lambda}$H and $\Omega$-hypernuclei.
The values of parameters $a$ and $b$ in $R_{f}(p_{T})$ for $H(p\Omega^-)$ and $H(n\Omega^-)$ are the same with that of the deuteron, and those for $^3_{\Lambda}$H and $H(pn\Omega^-)$ are the same with $^3$He.
So our calculated results for the $^3_{\Lambda}$H and $\Omega$-hypernuclei are parameter-free, and they are more potent for further test of the coalescence mechanism in describing the productions of nuclei with strangeness flavor quantum number.

\subsection{The results of the $^3_{\Lambda}$H} 

\begin{figure}[!htb]
\includegraphics[width=0.9\hsize]{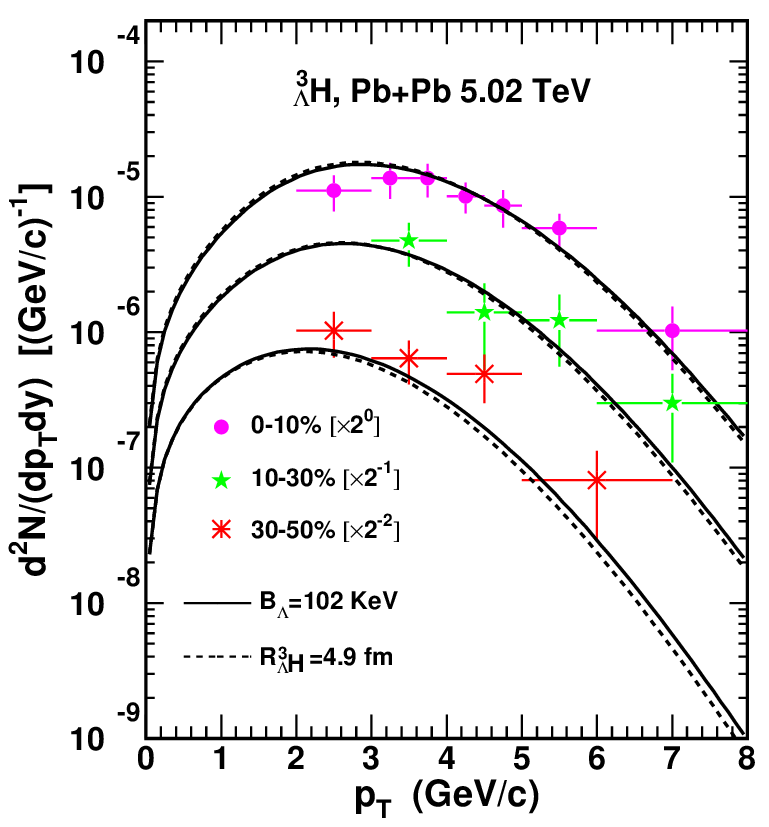}
\caption{The $p_T$ spectra of the $^{3}_{\Lambda}$H in Pb+Pb collisions at $\sqrt{s_{NN}}=5.02$ TeV. 
Filled symbols with error bars are the experimental data~\cite{ALICE:2024koa}.
The solid and dashed lines are the theoretical results with a halo structure and a spherical shape, respectively.}
\label{fig:hypertritonpt}
\end{figure}

Based on Eq.~(\ref{eq:pt-Hj3h}), we in this subsection compute the production of the $^{3}_{\Lambda}$H. 
Considering that the experimental measurements of the $^{3}_{\Lambda}$H suggest a halo structure with a $d$ core encircled by a $\Lambda$,
we first use $\sigma_1=\sqrt{\frac{2(m_p+m_n)^2}{3(m_p^2+m_n^2)}} R_d$ fm and $\sigma_2=\sqrt{\frac{2(m_d+m_{\Lambda})^2}{9(m_d^2+m_{\Lambda}^2)}} r_{\Lambda d}$.
The $\Lambda-d$ distance $r_{\Lambda d}$ is evaluated via $r_{\Lambda d}=\sqrt{\hbar^2/(4\mu B_{\Lambda})}$~\cite{Bertulani:2022vad},
where $\mu$ is the reduced mass and the binding energy $B_{\Lambda}$ here is adopted to be the latest and most precise measurement to date 102 KeV~\cite{ALICE:2022sco}.
We also take a spherical shape for the $^{3}_{\Lambda}$H to execute the calculation to study the influence of the shape on its production. 
In this case, $\sigma_1=\sqrt{\frac{m_{\Lambda}(m_p+m_n)(m_p+m_n+m_{\Lambda})} {m_pm_n(m_p+m_n)+m_nm_{\Lambda}(m_n+m_{\Lambda})+m_{\Lambda}m_p(m_{\Lambda}+m_p)}} R_{^{3}_{\Lambda}\text{H}} $,
and $\sigma_2=\sqrt{\frac{4m_pm_n(m_p+m_n+m_{\Lambda})^2} {3(m_p+m_n)[m_pm_n(m_p+m_n)+m_nm_{\Lambda}(m_n+m_{\Lambda})+m_{\Lambda}m_p(m_{\Lambda}+m_p)]}} R_{^{3}_{\Lambda}\text{H}}$,
where the root-mean-square radius $R_{^{3}_{\Lambda}\text{H}}$ is adopted to be 4.9 fm~\cite{Nemura:1999qp}.
Fig.~\ref{fig:hypertritonpt} shows the $p_T$ spectra of the $^{3}_{\Lambda}$H in 0-10\%, 10-30\% and 30-50\% centralities in Pb+Pb collisions at $\sqrt{s_{NN}}=5.02$ TeV. 
Filled symbols with error bars are the experimental data~\cite{ALICE:2024koa}.
The solid lines are the theoretical results of the coalescence model with a halo structure and the dashed lines are those for a spherical shape.
The $p_T$ spectra in different centralities are scaled by different factors for clarity as shown in the figure.
From Fig.~\ref{fig:hypertritonpt}, one can see that there exists a weak difference in the theoretical results of the $p_T$ spectra between a halo structure and a spherical shape, and the latter gives a little softer $p_T$ spectra. 
The results with a halo structure approach to the available data better, both for amplitude and for shape.
This point can also be seen in the results of the averaged transverse momenta $\langle p_T\rangle$ and yield rapidity densities $dN/dy$ of $^3_{\Lambda}$H hereunder.

\begin{table*}[!htb]
\begin{center}
\caption{Averaged transverse momenta $\langle p_T\rangle$ and yield rapidity densities $dN/dy$ of $^3_{\Lambda}$H in different centralities in Pb+Pb collisions at $\sqrt{s_{NN}}=5.02$ TeV. Experimental data in the seventh column are from Ref.~\cite{ALICE:2024koa}. Theory-4.9 denotes theoretical results with a spherical shape at $R_{^{3}_{\Lambda}\text{H}}=4.9$ fm. Theory-102, Theory-148 and Theory-410 denote theoretical results with a halo structure at $B_{\Lambda}=$102, 148 and 410 KeV, respectively.}  \label{table:dNdypt-hyt}
\begin{tabular}{@{\extracolsep{\fill}}ccccccccccccc@{\extracolsep{\fill}}}
\toprule
     &\multirow{2}{*}{Centrality}     &\multicolumn{4}{@{}c@{}}{$\langle p_T\rangle$ (GeV)}          &          &\multicolumn{5}{@{}c@{}}{$dN/dy~~(\times10^{-5})$} \\
\cline{3-6}  \cline{8-12} 
                   &     &Theory-4.9     &Theory-102   &Theory-148   &Theory-410                 &          &Data     &Theory-4.9       &Theory-102   &Theory-148  &Theory-410  \\
\hline
\multirow{3}{*}{$^3_{\Lambda}$H}
 &0-10\%             &$3.16$      &$3.19$    &$3.24$   &$3.37$                 &          &$4.83\pm0.23\pm0.57$  &$6.09$  &$5.96$  &$7.75$  &$12.7$    \\
 &10-30\%            &$2.90$      &$2.94$    &$2.99$   &$3.11$                &          &$2.62\pm0.25\pm0.40$  &$2.98$   &$2.99$  &$4.07$  &$7.44$    \\
 &30-50\%            &$2.46$     &$2.52$    &$2.55$   &$2.65$               &          &$1.27\pm0.10\pm0.14$   &$0.875$ &$0.932$  &$1.35$ &$2.94$    \\
\botrule
\end{tabular}
\end{center}
\end{table*}

Table \ref{table:dNdypt-hyt} presents the averaged transverse momenta $\langle p_T\rangle$ and yield rapidity densities $dN/dy$ of $^3_{\Lambda}$H in different centralities in Pb+Pb collisions at $\sqrt{s_{NN}}=5.02$ TeV.
Experimental data in the seventh column are from Ref.~\cite{ALICE:2024koa}.
Theory-4.9 in the third and eighth columns denotes theoretical results with a spherical shape at $R_{^{3}_{\Lambda}\text{H}}=4.9$ fm.
Theory-102 in the fourth and ninth columns are theoretical results at $B_{\Lambda}=$102 KeV.
Theory-148 in the fifth and tenth columns are theoretical results at a word averaged value of $B_{\Lambda}=$148 KeV~\cite{ALICE:2022sco}.
We also give theoretical results at $B_{\Lambda}=$410 KeV measured by the STAR collaboration~\cite{STAR:2019wjm} in the sixth and eleventh columns.
A clear decreasing trend for both $\langle p_T\rangle$ and $dN/dy$ from central to semi-central collisions is observed.
This is the same as light nuclei, which is due to that in more central collisions more energy is deposited in the midrapidity region and collective evolution exists longer.
For the halo structure, with the increase of the $B_{\Lambda}$, the size of the $^3_{\Lambda}$H decreases, and the suppression effect from the $^3_{\Lambda}$H size becomes relatively weak.
This leads to an increase for $dN/dy$ with the increasing $B_{\Lambda}$.
Besides $dN/dy$, such production suppression effect also affects the $p_T$ distribution~\cite{Wang:2023rpd,Liu:2024ygk}.
This is because the suppression effect becomes stronger with larger nucleus size in smaller system.
Recalling that $R_f(p_T)$ decreases with $p_T$, the $^3_{\Lambda}$H production is more suppressed in larger $p_T$ areas in the case of larger $^3_{\Lambda}$H size.
So there exists a decreasing trend for $\langle p_T\rangle$ with the decreasing $B_{\Lambda}$, as shown in Table \ref{table:dNdypt-hyt}.
This is the reason why the $\langle p_T\rangle$ of $^3_{\Lambda}$H is even smaller than that of the triton while the $\langle p_T\rangle$ of $\Lambda$ is larger than the nucleon.

\subsection{Predictions of $\Omega$-hypernuclei} 

\begin{figure*}[!htb]
\includegraphics[width=0.8\hsize]{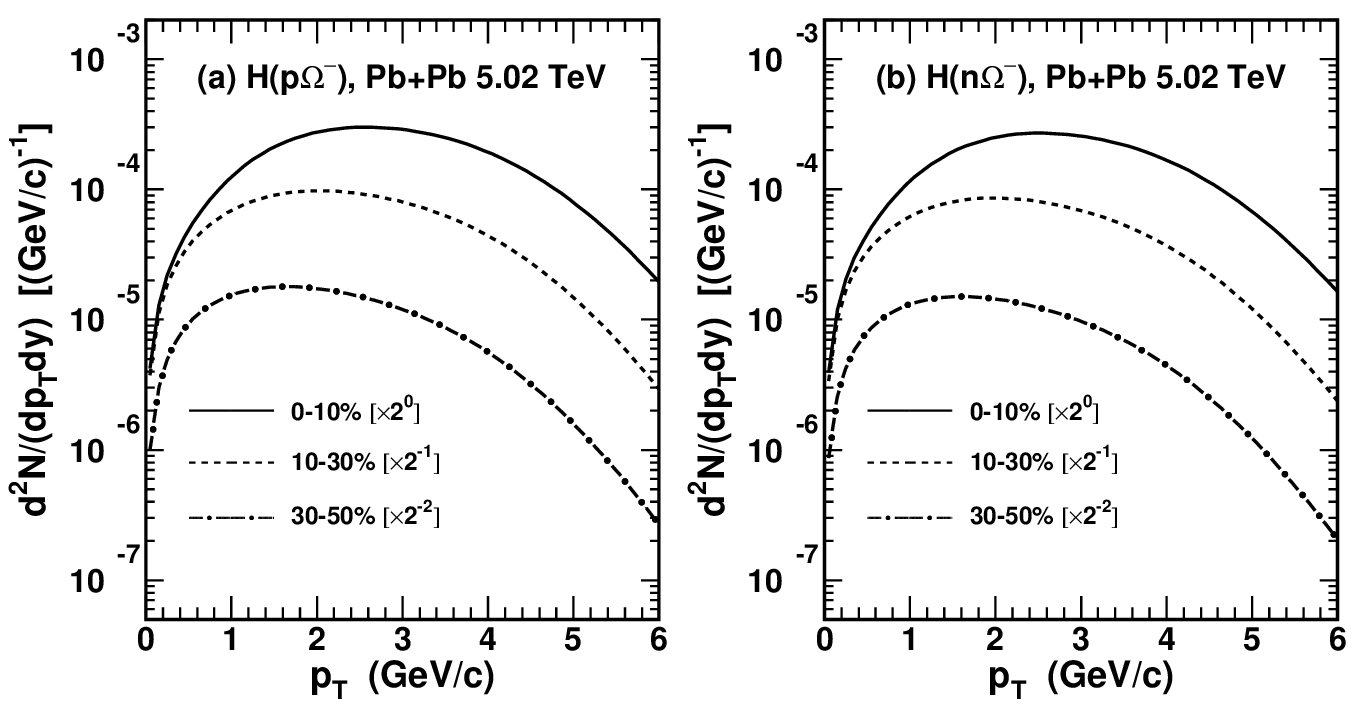}
\caption{Predictions of the $p_T$ spectra of (a) $H(p\Omega^-)$ and (b) $H(n\Omega^-)$ in different centralities in Pb+Pb collisions at $\sqrt{s_{NN}}=5.02$ TeV. }
\label{fig:pOnOpt}
\end{figure*}

The nucleon-$\Omega$ dibaryon in the S-wave and spin-2 channel is an interesting candidate of the deuteron-like state~\cite{Clement:2016vnl}.
The HAL QCD collaboration has reported the root mean square radius of $H(p\Omega^-)$ is about 3.24 fm and that of $H(n\Omega^-)$ is 3.77 fm~\cite{HALQCD:2018qyu}.
According to Eq.~(\ref{eq:pt-Hj2h}), we study the productions of $H(p\Omega^-)$ and $H(n\Omega^-)$, where the spin degeneracy factor $g_{H(p\Omega^-)}=g_{H(p\Omega^-)}=5/8$.
Fig.~\ref{fig:pOnOpt} shows predictions for their $p_T$ spectra in 0-10\%, 10-30\% and 30-50\% centralities with solid, dashed and dash-dotted lines, respectively, in Pb+Pb collisions at $\sqrt{s_{NN}}=5.02$ TeV. 
Different lines are scaled by different factors for clarity as shown in the figure.

\begin{table}[!htb]
\begin{center}
\caption{Predictions of averaged transverse momenta $\langle p_T\rangle$ and yield rapidity densities $dN/dy$ of $H(p\Omega^-)$ and $H(n\Omega^-)$ in different centralities in Pb+Pb collisions at $\sqrt{s_{NN}}=5.02$ TeV.}  \label{table:dNdypt-pOnO}
\begin{tabular}{@{\extracolsep{\fill}}cccc@{\extracolsep{\fill}}}
\toprule
     &Centrality     &$~~\langle p_T\rangle$ (GeV)          &$~~dN/dy~~(\times10^{-4})$ \\
\hline
\multirow{3}{*}{$H(p\Omega^-)~~$}
 &0-10\%                   &$2.84$    &$9.80$     \\
 &10-30\%                 &$2.44$    &$6.27$   \\
 &30-50\%                 &$2.18$    &$2.16$     \\
\hline
\multirow{3}{*}{$H(n\Omega^-)~~$}
 &0-10\%                   &$2.81$    &$8.75$     \\
 &10-30\%                 &$2.41$    &$5.46$   \\
 &30-50\%                 &$2.15$    &$1.79$     \\
\botrule
\end{tabular}
\end{center}
\end{table}

Table \ref{table:dNdypt-pOnO} presents predictions of the averaged transverse momenta $\langle p_T\rangle$ and yield rapidity densities $dN/dy$ of $H(p\Omega^-)$ and $H(n\Omega^-)$.
Both of them decrease from central to semi-central collisions, similar as light nucei and the $^3_{\Lambda}$H.
The very slight low results of $H(n\Omega^-)$ than $H(p\Omega^-)$ come from its slightly larger size.
Our predictions for $dN/dy$ of $H(p\Omega^-)$ and $H(n\Omega^-)$ are in the same magnitude with BLWC and AMPTC models in Ref.~\cite{Zhang:2020dma}.

\begin{figure}[!htb]
\includegraphics[width=0.8\hsize]{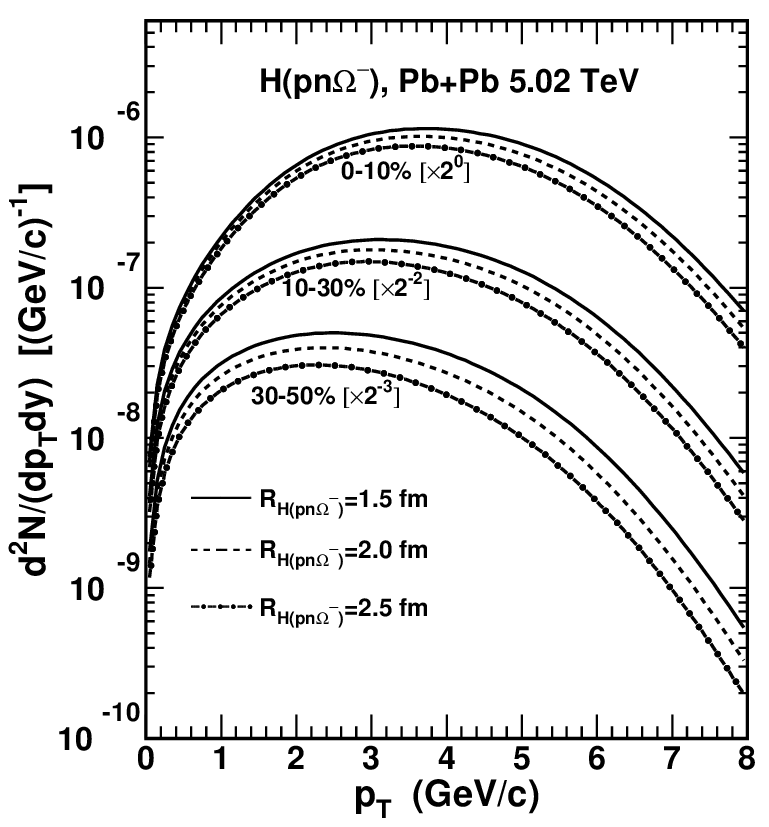}
\caption{Predictions of the $p_T$ spectra of $H(pn\Omega^-)$ in different centralities in Pb+Pb collisions at $\sqrt{s_{NN}}=5.02$ TeV. }
\label{fig:pnOpt}
\end{figure}

The $H(pn\Omega^-)$ with maximal spin-$\frac{5}{2}$ is proposed to be one of the most promising partners of the $t$ and $^3_{\Lambda}$H with multi-strangeness flavor quantum number~\cite{Garcilazo:2019igo}.
With Eq.~(\ref{eq:pt-Hj3h}), we study its production, and the spin degeneracy factor $g_{H(pn\Omega^-)}=3/8$.
As its root-mean-square radius $R_{H(pn\Omega^-)}$ is undetermined, we adopt 1.5, 2.0 and 2.5 fm to execute calculations, respectively.
Fig.~\ref{fig:pnOpt} shows predictions of the $p_T$ spectra in 0-10\%, 10-30\% and 30-50\% centralities in Pb+Pb collisions at $\sqrt{s_{NN}}=5.02$ TeV.
Solid, dashed and dash-dotted lines denote results with $R_{H(pn\Omega^-)}=$1.5, 2.0 and 2.5 fm, respectively, which are scaled by different factors for clarity as shown in the figure.

\begin{table*}[!htb]
\begin{center}
\caption{Predictions of averaged transverse momenta $\langle p_T\rangle$ and yield rapidity densities $dN/dy$ of $H(pn\Omega^-)$ in different centralities in Pb+Pb collisions at $\sqrt{s_{NN}}=5.02$ TeV. Theory-1.5, Theory-2.0 and Theory-2.5 denote theoretical results at $R_{H(pn\Omega^-)}=$1.5, 2.0 and 2.5 fm, respectively.}  \label{table:dNdypt-pnO}
\begin{tabular}{@{\extracolsep{\fill}}cccccccccc@{\extracolsep{\fill}}}
\toprule
     &\multirow{2}{*}{Centrality}     &\multicolumn{3}{@{}c@{}}{$\langle p_T\rangle$ (GeV)}          &          &\multicolumn{3}{@{}c@{}}{$dN/dy~~(\times10^{-6})$} \\
\cline{3-5}  \cline{7-9} 
                   &     &Theory-1.5   &Theory-2.0   &Theory-2.5                 &          &Theory-1.5       &Theory-2.0   &Theory-2.5    \\
\hline
\multirow{3}{*}{$H(pn\Omega^-)$}
 &0-10\%                   &$3.94$    &$3.88$   &$3.82$                 &          &$4.77$  &$4.17$  &$3.56$       \\
 &10-30\%                 &$3.44$    &$3.36$   &$3.29$                &          &$3.50$  &$2.95$  &$2.41$       \\
 &30-50\%                 &$2.98$    &$2.89$   &$2.81$               &          &$1.60$  &$1.24$  &$0.92$   \\
\botrule
\end{tabular}
\end{center}
\end{table*}

Table \ref{table:dNdypt-pnO} presents predictions of the averaged transverse momenta $\langle p_T\rangle$ and yield rapidity densities $dN/dy$ of $H(pn\Omega^-)$ in different centralities in Pb+Pb collisions at $\sqrt{s_{NN}}=5.02$ TeV.
Theory-1.5, Theory-2.0 and Theory-2.5 denote theoretical results at $R_{H(pn\Omega^-)}=$1.5, 2.0 and 2.5 fm, respectively.
Our predictions for $dN/dy$ are in the same magnitude with those in Ref.~\cite{Zhang:2021vsf}.

\subsection{Averaged transverse momentum ratios and yield ratios} \label{RptRyield}

\begin{figure*}[!htb]
\includegraphics[width=0.95\hsize]{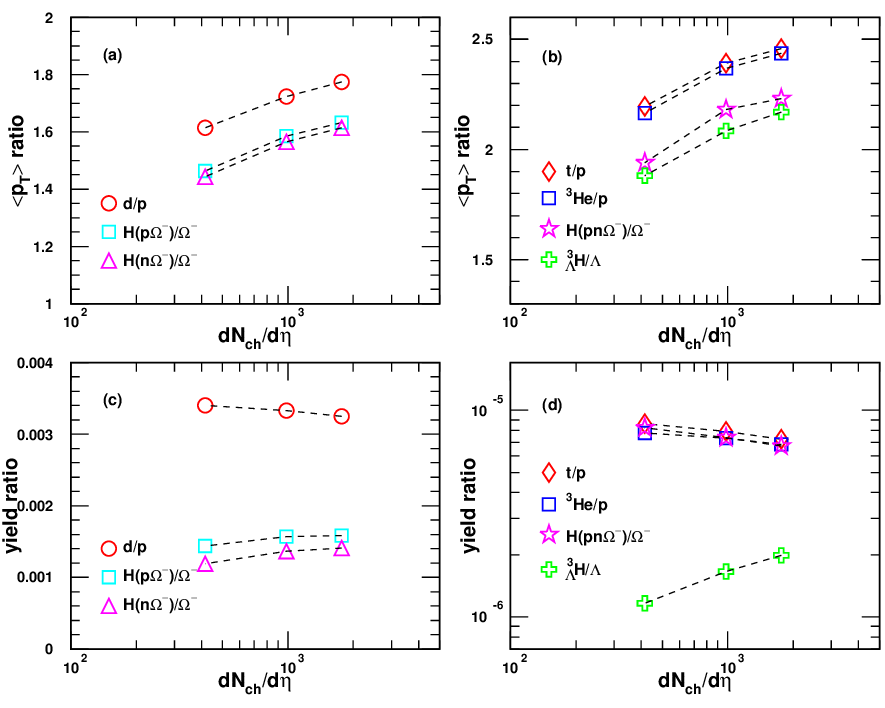}
\caption{The $\langle p_T\rangle$ ratios of (a) dibaryon states to baryons, (b) tribaryon states to baryons and the yield ratios of (c) dibaryon states to baryons, (d) tribaryon states to baryons as a function of $dN_{ch}/d\eta$ in Pb+Pb collisions at $\sqrt{s_{NN}}=5.02$ TeV. }
\label{fig:RaveptdNdy}
\end{figure*}

Based on the results of light nuclei and hypernuclei above, we find two groups of interesting observables which are powerful probes for the nucleus production mechanism.
One group of observables are the $\langle p_T\rangle$ ratios of light nuclei to protons and hypernuclei to hyperons.
The other are the centrality-dependent yield ratios of theirs.

Fig.~\ref{fig:RaveptdNdy} (a) and (b) show the $\langle p_T\rangle$ ratios of dibaryon states to baryons and those of tribaryon states to baryons, i.e.,
$\frac{\langle p_T\rangle_d}{\langle p_T\rangle_p}$, $\frac{\langle p_T\rangle_{H(p\Omega^-)}}{\langle p_T\rangle_{\Omega^-}}$, $\frac{\langle p_T\rangle_{H(n\Omega^-)}}{\langle p_T\rangle_{\Omega^-}}$, 
$\frac{\langle p_T\rangle_t}{\langle p_T\rangle_p}$, $\frac{\langle p_T\rangle_{\mathrm{^3He}}}{\langle p_T\rangle_p}$, $\frac{\langle p_T\rangle_{^3_{\Lambda}\mathrm{H}}}{\langle p_T\rangle_{\Lambda}}$ and $\frac{\langle p_T\rangle_{H(pn\Omega^-)}}{\langle p_T\rangle_{\Omega^-}}$. 
Open symbols connected by dashed lines to guide the eyes are the theoretical results of the coalescence model.
All these $\langle p_T\rangle$ ratios increase as a function of $dN_{ch}/d\eta$ due to the stronger collective flow in more central collisions.
More interestingly, these $\langle p_T\rangle$ ratios of light nuclei to protons and hypernuclei to hyperons happen to offset the $\langle p_T\rangle$ differences of $p$, $\Lambda$ and $\Omega^-$.
This makes them more powerful to bring characteristics resulted from the production mechanism to light.
Both dibaryon-to-baryon and tribaryon-to-baryon $\langle p_T\rangle$ ratios exhibit a distinct reverse-hierarchy of the light (hyper-)nucleus sizes at any centrality, i.e.,
$\frac{\langle p_T\rangle_d}{\langle p_T\rangle_p} > \frac{\langle p_T\rangle_{H(p\Omega^-)}}{\langle p_T\rangle_{\Omega^-}} > \frac{\langle p_T\rangle_{H(n\Omega^-)}}{\langle p_T\rangle_{\Omega^-}}$ as $R_d < R_{H(p\Omega^-)} < R_{H(n\Omega^-)}$,
and $\frac{\langle p_T\rangle_t}{\langle p_T\rangle_p} > \frac{\langle p_T\rangle_{\mathrm{^3He}}}{\langle p_T\rangle_p} > \frac{\langle p_T\rangle_{H(pn\Omega^-)}}{\langle p_T\rangle_{\Omega^-}} > \frac{\langle p_T\rangle_{^3_{\Lambda}\mathrm{H}}}{\langle p_T\rangle_{\Lambda}}$ as $R_t < R_{\mathrm{^3He}} < R_{H(pn\Omega^-)} < R_{^3_{\Lambda}\mathrm{H}}$. 
Here we take $R_{H(pn\Omega^-)}=2$ fm.
Such reverse-hierarchy comes from stronger production suppression for light (hyper-) nuclei with larger sizes in higher $p_T$ region.
This production property is very different from the thermal model in which these ratios are approximately equal to each other~\cite{Andronic:2010qu}.

Fig.~\ref{fig:RaveptdNdy} (c) and (d) show yield ratios of dibaryon states to baryons and those of tribaryon states to baryons.
Open symbols connected with dashed lines to guide the eye are the theoretical results of the coalescence model.
Some of these ratios such as $d/p$, $t/p$, $^3\mathrm{He}/p$ and $H(pn\Omega^-)/\Omega^-$ decrease 
while the others $H(p\Omega^-)/\Omega^-$, $H(n\Omega^-)/\Omega^-$ and $^3_{\Lambda}\mathrm{H}/\Lambda$ increase as a function of $dN_{ch}/d\eta$.
From Eqs.~(\ref{eq:pt-Hj2h}) and (\ref{eq:pt-Hj3h}), similar as Eq.~(\ref{eq:yRdp}), we approximately have
{\setlength\arraycolsep{0.2pt}
\begin{eqnarray}
             && \frac{d}{p} \sim \frac{H(p\Omega^-)}{\Omega^-}  \sim \frac{H(n\Omega^-)}{\Omega^-}   \nonumber \\ 
\propto && \frac{ N_p }{\langle R_f \rangle^3 \left(C_w +\frac{\sigma_{d,H(p\Omega^-),H(n\Omega^-)}^2}{\langle R_f \rangle^2}\right)  \sqrt{\frac{C_w}{\langle\gamma\rangle^2}+\frac{\sigma_{d,H(p\Omega^-),H(n\Omega^-)}^2}{\langle R_f \rangle^2}}} \nonumber \\ 
          =&& \frac{N_p}{\langle R_f \rangle^3/\langle\gamma\rangle}  \times \frac{ 1 }{ \left(C_w +\frac{\sigma_{d,H(p\Omega^-),H(n\Omega^-)}^2}{\langle R_f \rangle^2}\right)  \sqrt{C_w+\frac{\sigma_{d,H(p\Omega^-),H(n\Omega^-)}^2}{\langle R_f \rangle^2/\langle\gamma\rangle^2}}},~~~   \label{eq:yRdiBB}  
\end{eqnarray} }%
and
{\setlength\arraycolsep{0.2pt}
\begin{eqnarray}
            && \frac{t}{p}  \sim  \frac{^3\text{He}}{p}  \sim  \frac{^3_{\Lambda}\mathrm{H}}{\Lambda}  \sim  \frac{H(pn\Omega^-)}{\Omega^-}    \nonumber \\ 
\propto &&  \frac{ N_p^2  }
{\langle R_f \rangle^6 \left(C_w + \frac{\sigma_1^2}{\langle R_f \rangle^2}\right)  \sqrt{\frac{C_w}{\langle\gamma\rangle^2} +\frac{\sigma_1^2}{\langle R_f \rangle^2}}  }  \nonumber  \\
&& \times \frac{1}{  \left(\frac{4C_w}{3} + \frac{\sigma_2^2}{\langle R_f \rangle^2}\right)   \sqrt{\frac{4C_w}{3\langle\gamma\rangle^2}+\frac{\sigma_2^2}{\langle R_f \rangle^2}}}    \nonumber \\ 
        = && \left( \frac{N_p}{\langle R_f \rangle^3/\langle\gamma\rangle} \right)^2  \frac{ 1 }
{ \left(C_w + \frac{\sigma_1^2}{\langle R_f \rangle^2}\right)  \sqrt{C_w +\frac{\sigma_1^2}{\langle R_f \rangle^2/\langle\gamma\rangle^2}}    }  \nonumber  \\
&& \times \frac{1}{ \left(\frac{4C_w}{3} + \frac{\sigma_2^2}{\langle R_f \rangle^2}\right) \sqrt{\frac{4C_w}{3}+\frac{\sigma_2^2}{\langle R_f \rangle^2/\langle\gamma\rangle^2}}}.   \label{eq:yRtriBB}        
\end{eqnarray} }%
Eqs.~(\ref{eq:yRdiBB}) and (\ref{eq:yRtriBB}) show that centrality-dependent behaviors of these two-particle yield ratios closely relate with the nucleon number density $\frac{N_p}{\langle R_f \rangle^3/\langle\gamma\rangle}$ and the  production suppression effect via the relative size of nuclei to hadronic source systems $\frac{\sigma_i}{\langle R_f \rangle}~~(i=d,H(p\Omega^-),H(n\Omega^-),1,2$ and $\sigma_i \propto$ the corresponding nucleus size).

For the limit case of the nuclei with very small (negligible) sizes compared to the hadronic system scale, the $dN_{ch}/d\eta$-dependent behaviors of their yield ratios to baryons are completely determined by the nucleon number density.
For the general case, the item $\frac{\sigma_i}{\langle R_f \rangle}$ suppresses these ratios and such suppression becomes weaker in larger hadronic systems.
This makes these yield ratios increase from peripheral to central collisions, i.e., with the increasing $dN_{ch}/d\eta$. 
The larger nucleus size, the stronger increase as a function of $dN_{ch}/d\eta$.
The nucleon density decreases with increasing $dN_{ch}/d\eta$~\cite{ALICE:2019hno}, which makes these ratios decrease.
As the root-mean-square radii of $d$, $t$, $^3\mathrm{He}$ and $H(pn\Omega^-)$ are about or smaller than 2 fm, the decreasing nucleon density dominates the behaviors of their yield ratios to baryons.
But for $H(p\Omega^-)$, $H(n\Omega^-)$ and $^3_{\Lambda}\mathrm{H}$, their root-mean-square radii are larger than 3 fm, the production suppression effect from their sizes becomes dominant, which leads their yield ratios to baryons increase as a function of $dN_{ch}/d\eta$.
Such different centrality-dependent behaviors can help to justify the nucleus own sizes in future experiments for more light nuclei and hypernuclei. 

\section{Summary}  \label{summary}

We extended the analytical coalescence model previously developed for the productions of light nuclei to include the hyperon coalescence to study production characteristics of $d$, $^3\overline{\text{He}}$, $\bar t$, $^3_{\Lambda}\mathrm{H}$ and $\Omega$-hypernuclei.
We derived the formula of the momentum distribution of two baryons coalescing into dibaryon states and that of three baryons coalescing into tribaryon states.
The relationships of dibaryon states and tribaryon states with primordial baryons in momentum space in the laboratory frame were given.
The effects of the hadronic system scale and the nucleus own size on the nucleus production were also clearly presented.

We applied the extended coalescence model to Pb+Pb collisions at $\sqrt{s_{NN}}=5.02$ TeV.
We explained the available data of the $B_2$ and $B_3$, the $p_T$ spectra, averaged transverse momenta and yield rapidity densities of the $d$, $^3\overline{\text{He}}$, $\bar t$, and $^3_{\Lambda}\mathrm{H}$ measured by the ALICE collaboration.
We provided predictions of the $p_T$ spectra, averaged transverse momenta and yield rapidity densities of different $\Omega$-hypernuclei, e.g., $H(p\Omega^-)$, $H(n\Omega^-)$, and $H(pn\Omega^-)$, for future experimental measurements.

More interestingly, we found two groups of novel observables. One group refered to the averaged transverse momentum ratio 
$\frac{\langle p_T\rangle_d}{\langle p_T\rangle_p}$, $\frac{\langle p_T\rangle_{H(p\Omega^-)}}{\langle p_T\rangle_{\Omega^-}}$, $\frac{\langle p_T\rangle_{H(n\Omega^-)}}{\langle p_T\rangle_{\Omega^-}}$, 
$\frac{\langle p_T\rangle_t}{\langle p_T\rangle_p}$, $\frac{\langle p_T\rangle_{\mathrm{^3He}}}{\langle p_T\rangle_p}$, $\frac{\langle p_T\rangle_{^3_{\Lambda}\mathrm{H}}}{\langle p_T\rangle_{\Lambda}}$, $\frac{\langle p_T\rangle_{H(pn\Omega^-)}}{\langle p_T\rangle_{\Omega^-}}$. 
They exhibited a reverse-hierarchy according to the sizes of the nuclei themselves at any collision centrality.
The other group involved the centrality-dependent yield ratios $d/p$, $H(p\Omega^-)/\Omega^-$, $H(n\Omega^-)/\Omega^-$, $t/p$, $^3\mathrm{He}/p$, $^3_{\Lambda}\mathrm{H}/\Lambda$ and $H(pn\Omega^-)/\Omega^-$.
Some of these yield ratios $d/p$, $t/p$, $^3\mathrm{He}/p$ and $H(pn\Omega^-)/\Omega^-$ decreased while the others $H(p\Omega^-)/\Omega^-$, $H(n\Omega^-)/\Omega^-$ and $^3_{\Lambda}\mathrm{H}/\Lambda$ increased as a function of $dN_{ch}/d\eta$ in the same coalescence framework.
Such different trends were caused by different production suppression degrees from the nucleus size.
The behaviors of these two groups of ratios in the coalescence mechanism were different from the thermal model.
They were powerful probes of the nucleus production mechanism and can help judge the nucleus own sizes.

\section*{Acknowledgements}

This work was supported in part by the National Natural Science Foundation of China under Grants No. 12175115 and No. 12375074.

\bibliographystyle{apsrev4-1}
\bibliography{myref}

\end{document}